\documentclass[twocolumn,epj]{svjour}

\usepackage[space]{grffile}
\usepackage{amsfonts,hhline,color,slashed}
\usepackage{mathrsfs,amsmath,amssymb,graphicx}
\usepackage[T1]{fontenc}
\usepackage[utf8]{inputenc}
\usepackage[english]{babel}
\usepackage{comment}
\usepackage{epsfig,ulem}
\usepackage{multirow}
\usepackage{array}
\usepackage{cite}

\newcommand{\qbarq}{\langle \bar  q q \rangle}
\newcommand{\ubaru}{\langle \bar u u \rangle}
\newcommand{\mrm}{\mathrm}
\newcommand{\dd}{\mathrm{d}}

\newcommand{\beq}{\begin{eqnarray}}
\newcommand{\eeq}{\end{eqnarray}}
\usepackage{xcolor}

\newcommand{\mucep}{{\mu_{\textrm{CEP}}}}
\newcommand{\Tcep}{{T_{\textrm{CEP}}}}

\newcommand{\idest}{\textit{i.e.}~}
\newcommand{\param}{\Lambda,m_0,G}

\newcommand{\ccite}[1]{{\color{black} \cite{#1}}}

\newcommand{\reffig}[1]{{\color{black}\mbox{Fig. (\ref{#1})}}}
\newcommand{\reftab}[1]{{\color{black}\mbox{Tab. \ref{#1}}}}
\newcommand{\refeq}[1]{{\color{black}\mbox{Eq. (\ref{#1})}}}
\newcommand{\refeqs}[3]{{\color{black}\mbox{Eqs. (\ref{#1},\ref{#2},\ref{#3})}}}

\newcommand{\refapp}[1]{{\color{black}\mbox{App. (\ref{#1})}}}

\begin{document}
\title{Sensitivity of predictions in an effective model -- application
  to the chiral critical end point position in the Nambu--Jona-Lasinio model\thanks{
We would like to thanks Robin Jodon for useful discussions.  This work
was partially supported by Project CERN/FP/123620/2011 developed under
the initiative QREN financed by the UE/FEDER through the program
COMPETE -''Programa Operacional Factores de Competitividade''.
}
}
\titlerunning{Sensitivity of the CEP position in NJL}

\author{%
  Alexandre Biguet\inst{1} \and%
  Hubert Hansen\inst{1} \and%
  Pedro Costa\inst{2}\and%
  Pierre Borgnat\inst{3}\and%
  Timothée Brugière\inst{1}}
\institute{%
  Institut de Physique Nucl\'eaire de Lyon, CNRS/IN2P3,
  Universit\'e Claude Bernard de Lyon, 69622 Villeurbanne Cedex,
  France \and%
  Centro de F\'isica
  Computacional, Departamento de F\'isica, Universidade de Coimbra,
  P-3004-516 Coimbra, Portugal\and%
  Laboratoire de Physique, CNRS, l'École normale supérieure de
  Lyon, 46, allée d'Italie 69364 Lyon cedex 07 France
}
\authorrunning{A.Biguet, H.Hansen, P.Costa, P.Borgnat and T.Brugière}
\mail{a.biguet@ipnl.fr}

\abstract{%
  The measurement of the position of the chiral critical end point
  (CEP) in the QCD phase diagram is under debate. While it is possible
  to predict its position by using effective models specifically built
  to reproduce some of the features of the underlying theory (QCD),
  the quality of the predictions (\textit{e.g.}, the CEP position)
  obtained by such effective models, depends on whether solving the
  model equations constitute a well- or ill-posed inverse problem.
  Considering these predictions as being inverse problems provides
  tools to evaluate if the problem is ill-conditioned, meaning that
  infinitesimal variations of the inputs of the model can cause
  comparatively large variations of the predictions. If it is
  ill-conditioned, it has major consequences because of finite
  variations that could come from
  experimental and/or theoretical errors.\\
  In the following, we shall apply such a reasoning on the predictions
  of a particular Nambu--Jona-Lasinio model within the mean field +
  ring approximations, with special attention to the prediction of the
  chiral CEP position in the $(T-\mu)$ plane.  We find that the
  problem is ill-conditioned (\idest very sensitive to input
  variations) for the $T$-coordinate of the CEP, whereas, it is
  well-posed for the $\mu$-coordinate of the CEP. As a consequence,
  when the chiral condensate varies in a $10$ MeV range, $\mucep$
  varies far less. \\
  As an illustration to understand how problematic this could be, we
  show that the main consequence when taking into account finite
  variation of the inputs, is that the existence of the CEP itself
  cannot be predicted anymore: for a deviation as low as 0.6 \% with
  respect to vacuum phenomenology (well within the estimation of the
  first correction to the ring approximation) the CEP may or may not
  exist.
}

\maketitle

%
%

\section{Introduction}

The critical end point (CEP) was proposed at the end of the eighties
\ccite{Asakawa:1989bq} and is still a very important subject of
discussion nowadays: at finite temperature and chemical potential the
most common phase diagram shows a first-order chiral phase transition
separating the hadronic phase from the quark phase; this first-order line
finishes at the CEP where the phase transition is of second-order and,
as $T$ increases and $\mu$ decreases, the phase transition becomes a
crossover.

The existence of the CEP is still an open problem for theoretical
studies based on QCD while its experimental search is in progress
\ccite{Abelev:2009bw,Aggarwal:2010cw,Tarnowsky:2011vk,Lacey:2014wqa,Akiba:2015jwa,
CPOD,Luo:2009sx,Gavai:2011sr,Gazdzicki:2011fx}.  

Due to its relevance for the QCD phase diagram the search for the CEP
becomes one important issue for the heavy ion collision (HIC) program
\ccite{CPOD,Luo:2009sx,Gavai:2011sr}: the search of the CEP and the
deconfinement transition \ccite{CPOD,Luo:2009sx,Gavai:2011sr} is being
undertaken in SPS at CERN \ccite{Gazdzicki:2011fx}, in RHIC at BNL
\ccite{Abelev:2009bw,Aggarwal:2010cw,Tarnowsky:2011vk}, and in the future facilities
FAIR at GSI and NICA at JIRN \ccite{NICA}.  The eventual confirmation
of the CEP existence would be one of the first QCD-like observables in
the medium to be discovered with important implications on the
constraint of several effective models.

From the theoretical point of view the existence of the CEP is not
consensual: even if older results from lattice QCD \ccite{Fodor:2004nz}
predict the existence of the CEP, once the transition is a crossover
at vanishing chemical potential, $\mu=0$,
\ccite{Borsanyi:2010cj,Bazavov:2011nk} it is possible that it may
remain of this type also at $\mu\ne0$.  Most of the effective models
like the Nambu--Jona-Lasinio (NJL) and Polyakov-loop
Nambu--Jona-Lasinio (PNJL) models
\ccite{Fukushima:2004,Ratti:2006,Costa:2007ie,Fukushima:2008b,Kashiwa:2008a,%
  Rossner:2008,Costa:2008gr,PNJLPCHH1,Costa2010}
and the Polyakov-loop-improved quark-meson (PQM) model
\ccite{Schaefer:2007pw,Herbst:2010rf} also present a first order chiral
phase transition that ends at the CEP.  However, each model has its
own value for the location of the CEP that depends, for example, on
the chosen parametrization \ccite{PNJLPCHH1}, on the strength of the
vector meson coupling and on the anomaly strength through the 't Hooft
coupling constant \ccite{Fukushima:2008wg}, as well as on the
Polyakov-loop. This led to several attempts to constrain some models
in order to understand if the CEP exists or not
\ccite{Bratovic:2012qs,Contrera:2012wj,Hell:2012da} namely by fixing
the vector meson coupling so that the slope of the pseudo-critical
temperature obtained in lattice QCD simulations
\ccite{Kaczmarek:2011zz} at small $\mu$ is reproduced
\ccite{Bratovic:2012qs}.  Another possible constraint is by showing the
existence of a first order QCD phase transition in compact star
interiors which lead to the evidence of a first order transition that
would prove the existence of at least one CEP in the QCD phase diagram
\ccite{Blaschke:2013ana}. 

To better understand the physical mechanism that generates a CEP, one
usual approach is to vary some parameters of an effective model to see
if the physics controlled by those parameters is relevant to the CEP
position prediction. When doing such studies (\textit{e.g.} \ccite{PNJLPCHH1})
we realize the need to have a more systematic and a more quantitative
way to proceed. We also realized that instead of varying parameters
independently it would be better to vary them in such a way that the
inputs phenomenology (used to constrain the parameter in the first
place) remains almost constant.  
\\
In this paper we develop a method to study some aspects of the
sensitivity of a specific NJL model we are studying, that relies on
methods used for inverse problems \ccite{Tarantola,Ramm:1997}.  Simply
put, an inverse problem consists in finding the model parametrization
best reproducing some input data. This can be achieved, for example,
by minimizing some merit function as a $\chi^2$ or --as it is the case
here-- by an exact inversion of the problem. Still, considering it as
an inverse problem is by far richer, because it is a framework aiming
at extracting and characterizing as much information as possible from
the data and their modelization \ccite{Tarantola}.  Note that inverse
problem is currently considered as a nice framework of study in the
nuclear physics community (\textit{e.g.}
\ccite{Dobaczewski1,Dobaczewski2,Dobaczewski1,Reinhard:2010,Fattoyev:2011}).
Especially, the guidelines described in \ccite{Dobaczewski1} were
particularly interesting and inspiring to develop our work.

For effective models such as the one we will study, related works have
shown some deep consequences of the idea that one is solving an
inverse problem.
In \ccite{Fattoyev:2011}, with a framework different from the one
presented here, it is shown that varying individually the model
parameters may be ``misleading and ill advised'' to determine the
uniqueness of a model if the inverse problem is not well posed.
In \ccite{Reinhard:2010}, the relevance of a systematic analysis of
the parameter space is discussed.
These both works \ccite{Reinhard:2010,Fattoyev:2011} stress that, if
the value of the merit function ($\chi^2$ therein) at minimum is a
measure of how well the resulting parametrization is able to
reproduced the input data, there are other meaningful information to
obtain.  The curvature around this minimum is one, assessing if the
model gives stable and meaningful prediction.  Therefore, it is
important not to get ``trapped in the $\chi^2$ minimum'' and to study
variations around the minimum so as to get access to the speed at
which the $\chi^2$ value deteriorate.

After the parameters fixing, one central question in using an
effective model is to estimate if it is reasonable to extrapolate it
away of the region where parameters has been fixed, if it keeps its
predictive power and how far.  For this work, an effective model means
the model Lagrangian with the addition of the input parameters, its
approximations and the way the parameters are computed.  For example,
if some in-vacuum inputs are used, one usually assumes that meaningful
results can be get at finite temperature (it is indeed one of the
earliest successes of the NJL model to show that the quark condensate
melts with
temperature); yet this remains to be evaluated. \\
In the present work, a sensitivity parameter is introduced as a way to
qualitatively estimate if a prediction is very sensitive to the input
(hence there is strong possibility that the predicted value cannot be
trusted) or if there is a reason to believe that the prediction is
well constrained by the model calculations and the chosen inputs. We
will say that in the former case the prediction is unstable (against
variation of the inputs) and
in the latter that the prediction is stable. \\
This type of sensitivity analysis is quite common in nuclear theory
\ccite{Kortelainen2010,Dobaczewski1}. However, the precise definition
of a sensitivity parameter varies from work to work, although they all
estimate how stable a model prediction is. Here, the inverse problem
that will be considered has an exact solution (if we where to define a
$\chi^2$, its minimum value would be zero) whereas in the
aforementioned work the parameters fit is not exact ($\chi^2 \neq
0$). More precisely we will define a sensitivity parameter that
measure how an infinitesimal variation of the inputs of
the model will impact the value of a prediction. \\
We found out that the sensitivity parameter that we define later on is
very closely related to a criterion defined in the computer science
community to estimate if the result of a numerical computation will be
damage because real numbers has to be approximated as float numbers
(propagation of round-off error) \ccite{FelipeCucker2002}. The
sensitivity parameter is indeed related to the so called condition
number \ccite{Demmel:1987} and is an estimation of the distance
between the problem at hand (for example the computation of the
solution of a linear system) and the closest ill-posed problem (in
that example it would be a non-invertible system). When the condition
number is large, the problem is said to be ill-conditioned in the
sense that small
errors in the data will results in large error in the outcomes. 
 
The proposed reasoning with inverse problems and the associated tool
may be quite involved.  Since, up to our knowledge, they have not been
used widely in the study of the phase of QCD, we will present here a
simple analysis based only on the sensitivity parameter and a
correlation analysis that was inspired among other works by
\ccite{Dobaczewski1}. We choose the SU(2) NJL model with interaction
in the scalar channel only, at the mean field + ring
approximation. The inputs will be the quark condensate $\qbarq$, the
pion mass $m_\pi$ and its decay constant $f_\pi$. We choose this model
because it is good enough to reproduce basic chiral properties of QCD
(dynamical mass generation via the spontaneous breaking of chiral
symmetry and a possible CEP) but simple enough to be able to exactly
solve the inverse problem and doing part of the calculation
semi-analytically. The simplicity of the model and its strong symmetry
properties enable us to concentrate on discussing the usefulness of
this analysis, to better understand the role of the sensitivity and
check the validity of our computation.  Even with such simplification
we obtain useful results on one key
observable of the QCD phase diagram, namely the CEP position. \\

\noindent
The paper is organized as follow: \\
In the first part we will introduce the NJL model and quickly review its
relevant phenomenology for this work (spontaneous chiral symmetry
breaking, pion properties and the chiral critical end point). \\
Then we will define the sensitivity parameter, compute its value in
the NJL model for several predictions (the sigma meson mass and the
pion-quark-antiquark effective coupling constant in vacuum together
with the position of the chiral CEP in the
$(T-\mu)$ plane) and discuss its relevance to characterize the well- and
ill-posedness of the problem. \\
We will also present an analysis of the situation when one relaxes one
of the constraint of the model, namely the value of the quark
condensate. 

In the second part, the consequences of the previous study will be
discussed when finite variation of the fitting data, $m_\pi$, $f_\pi$
and $\qbarq$ are considered. It will illustrate that if the
sensitivity analysis may seems fairly abstract, for low value of the
dispersion that are well within the expected range of the correction
generated by using a next to leading order approximation, the physics
can drastically change (namely the CEP may disappear).

%
%

\section{Sensitivity of predictions of the Nambu--Jona-Lasinio model}

\subsection{Parametrization of the Nambu--Jona-Lasinio model;
  observables}

We consider the local two flavor NJL model in $SU(2)$-isospin
approximation whose Lagrangian is (see
\ccite{NJLrev_klevansky,NJLrev_weise, NJLrev_buballa,NJLrev_hatsuda}
for reviews):
\begin{equation}
  \label{eq:NJL-Lagrangian}
  \mathcal{L}_{NJL} = \bar{\psi} ( i\gamma^\mu \partial_\mu - m_0 ) \psi 
                 + G \left[ (\bar{\psi}\psi)^2  
                 + ( \bar\psi i \gamma_5 \boldsymbol\tau \psi )^2 \right] \;. 
\end{equation}
This Lagrangian depends on three dimensional parameters, $m_0$ the
bare quark mass of the $u$ and $d$ quarks in SU(2)-isospin
approximation (in GeV); $G$ the coupling constant (GeV$^{-2}$) and
$\Lambda$ the three-dimensional cutoff mimicking the asymptotic freedom of
quarks (in GeV). Since the NJL model cannot be extracted directly from QCD, they
are free parameters, but, nevertheless, they are loosely constrained:
$m_0$ should be of the order of the masses of $u$ and $d$ quarks, $\Lambda$
is related to the scale $\Lambda_{QCD}$; for what concern $G$ if one think
of it as the Fermi coupling in the electroweak theory, $G \simeq g/M^2
\simeq \tilde g / \Lambda^2$, then $\tilde g$ is poorly constrained but is
usually expected to be in a range $[1, 10]$.

These parameters are usually fitted to the values of the pion mass, $m_\pi$,
the pion decay constant $f_\pi$, and the quark condensate $c =
-\qbarq^{1/3}$ normalized to be positive and with the dimension of an
energy. The latter quantity is related to the so-called dressed quark
mass $m$ that cannot be considered as an observable but provides a
physical picture of the hadronic world after the spontaneous chiral
symmetry breaking in terms of quasi-particles with $m \gg m_0$ and
also a very crude approximation of the nucleon mass as 3 times the
mass of one dressed quark. \\

The scheme we choose to compute those quantities is the mean field
approximation for the condensate and the ring approximation for the
meson properties\ccite{NJLrev_klevansky}. \\
For the mean field effective quark mass $m(\param)$ one has the so-called
gap equation:
\begin{equation}
  \label{eq:gap-equation}
  m_0 - m + 8iGN_cN_f m I_1 = 0 \;
\end{equation}
where $I_1$ is the 1-propagator line integral that arises from the
tadpole diagram:
\beq
\label{eq:I1}
I_1 &=& -i \int^\Lambda \frac{\dd^3 p }{(2\pi)^3} \frac{1}{2E_p} \mbox{
  (with $E_p^2 = p^2 + m^2$)}.
\eeq
Then the mean field quark condensate \mbox{$\qbarq (\param)$} is:
\beq
\label{eq:cond}
    \qbarq   & = & \frac{m_0 - m}{2G}\;.
\eeq
Finally $m_\pi(\param)$ and $f_\pi(\param)$ are given by:
\beq
\label{eq:mpi}
    m_\pi^2   & = &  - \frac{m_0}{m} \frac{1}{4 i G N_c N_f I_2(0)} \;, \\
\label{eq:fpi}
    f_\pi^2   & = &  -4i N_c m^2 I_2(0)  \;,
\eeq
where $I_2$ is the 2-propagator lines integral coming from the quark loop diagram of the ring
approximation ; we also use a quasi-Goldstone boson
approximation assuming the pion mass can be neglected, namely the
argument of $I_2$ is $k^2 = 0$ and not $k^2 = m_\pi^2$. Explicitly one
has:
\beq
\label{eq:I2}
I_2(0) &=& -i \int^\Lambda \frac{\dd^3 p }{(2\pi)^3} \frac{1}{4E_p^3}.
\eeq
\medskip

When $(m_\pi,\ f_\pi, \qbarq)$ are fixed to their phenomenological
values, the inverse problem is solved when the system:
\begin{align}
\label{eq:inverse-problem-sys-njl-1}
    m_\pi    (\param) & =  m_\pi  \;, \\
\label{eq:inverse-problem-sys-njl-2}
    f_\pi    (\param) & =  f_\pi  \;,   \\
\label{eq:inverse-problem-sys-njl-3}
\qbarq (\param) & =  \qbarq \;, 
\end{align}
is solved for the parameters $\Lambda$, $m_0$ and $G$. Thanks to the
quasi-Goldstone approximation this system has quasi-analytic
solutions discussed in \refapp{app:inverse-pb}. As we show in this
appendix, the previous system has solutions only if the ratio $\alpha =
f_\pi^3 / \qbarq$ is greater than a constant critical value (it is
related to the discussion in Sec. 2.2.2 of
\ccite{NJLrev_buballa}). Among the solutions, only one is physical.

When parameters are fixed, this NJL model can describe some simple 
phenomenology in vacuum as the sigma meson mass $m_\sigma$ and the pion-quark
coupling constant $g_{\pi \bar q q}$ (see\ccite{NJLrev_klevansky} for details):
\beq
  m_\sigma &=& \sqrt{4 m^2 + m_\pi^2} \;, \nonumber \\
  g_{\pi \bar q q} &=& \frac{1}{\sqrt{-4i N_c I_2(0)}} \;.
\label{eq:vac-predictions}
\eeq

The NJL model is also able to predict, for a range of parameters, a
first order transition toward a phase where the chiral symmetry
is partially restored. At the end of this line there is a critical end
point in the $(T-\mu)$ plane whose properties are described in
\refapp{app:cep-calc} (we also describe a new very fast and stable
algorithm to compute it).

\subsection{Sensitivity and ill-posedness of a problem}

To study the sensitivity of a given prediction we will use a condition
number \ccite{Demmel:1987} of the problem: it is a local measure
(based on a gradient calculation) of the sensitivity of a solution to
this problem against infinitesimal variation of its inputs. Here we
will use the relative condition number: when it is infinite, the
problem is ill-posed; when it is finite but large
(compared to one) the problem is said ill-conditioned. \\
The choice we made to compute it is based on the statistical
propagation of errors because it is a natural way to compute the variation
(the standard deviation) of an output with respect to input
variations that are supposed uncorrelated when one minimizing a $\chi^2$
(as we will do in the future where an exact inversion is not possible)
and for the correlation analysis. \\
Let $X$ be a prediction depending on two inputs $a$ and $b$: the
standard deviation of a prediction $X$ is computed by propagating the
variation $\sigma(a)$ and $\sigma(b)$ of the parameters:
\begin{equation}
\sigma^2(X) = \left(\frac{\partial X}{\partial a}\right)^2 \sigma^2(a)
+ \left(\frac{\partial X}{\partial b}\right)^2 \sigma^2(b) \;.
\end{equation}
The sensitivity is the ratio of the relative standard deviation
and the mean of the relative variation of the inputs:
\begin{equation}
\Sigma(X) = \lim_{\sigma \rightarrow 0} \frac{\sigma_{rel}(X)} {\sigma_{rel}^{I}}
\end{equation}
where,
\beq
  \sigma_{rel}(X) &=& \frac{\sigma(X)} X \\
  \sigma_{rel}^{I} &=& \frac 1 2 \left( \frac{\sigma(a)} a + \frac{\sigma(b)} b \right)\,,
\eeq
and $\lim_{\sigma \rightarrow 0}$ means we take infinitesimal variations
of the inputs. The way one takes this limit has to be specified. We
choose to take vanishing \textit{relative dispersion} of the inputs
namely for $I = a$ or $b$, $\sigma(I) / I = p$ and $p \rightarrow 0$.  We
will briefly discuss another choice (to cross check our results), the
vanishing \textit{absolute dispersion} case, where if all inputs have
the same dimension, $\sigma(I) = d$ and $d \rightarrow 0$.

Explicitly with the relative dispersion and for the inputs $(m_\pi,
f_\pi, c)$ we have:
\begin{equation}
  \label{eq:sensitivity}
  \Sigma(X) = \sqrt{
    \left(\frac{\partial X}{\partial m_\pi}\right)^2 \frac{m_\pi ^2}{X^2}
    + \left(\frac{\partial X}{\partial f_\pi}\right)^2 \frac{f_\pi ^2}{X^2}
    + \left(\frac{\partial X}{\partial c}\right)^2 \frac{c^2}{X^2}
  } 
\;,
\end{equation}
where it can be noticed that:
\begin{equation}
\left( \frac{\partial X}{\partial a} \frac{a}{X}\right)^2 = \left(
  \frac{\partial \ln X}{\partial \ln a} \right)^2 \;.
\end{equation}

In the rest of this paper we will argue that if the sensitivity of a
prediction is large (the calculation is ill-conditioned) it means that it
cannot be trusted and should be excluded because any small but finite
errors in the inputs (either measurement errors as for the condensate
or theoretical errors because of the approximations) will have a great
chance of damaging the prediction. We will say that the prediction is
unstable. \\
On the contrary, a prediction with a small sensitivity can be trusted
in the context of the particular model used to predict it.  The
physics that has been used to write the model and the chosen inputs
are enough to give a stable, well constrained prediction.

In our work, we choose to use a model describing the chiral physics
where spontaneous chiral symmetry breaking is generated only by the
scalar interaction and the inputs are related to this physics: the
pion properties (the quasi-Goldstone boson of this mechanism whose
mass and decay constant are related via the PCAC) and the non vanishing
chiral condensate.

We will illustrate the use of the sensitivity with 3 cases: stable
prediction in vacuum, stable and unstable prediction in medium at zero
and finite dispersion.

\subsubsection{A remark on the sensitivity calculation: Monte Carlo setup}

It is in principle possible to compute the sensitivity
analytically. In the present context this can actually be done quite
easily for the in-vacuum predictions because the inverse problem is
exact. Details of the calculation of $\Sigma(m_\sigma)$ can be found
in \refapp{app:msigma-sensitivity}. In the simple model we considered,
the calculation of $\Sigma(\Tcep)$ and $\Sigma(\mucep)$ is still
treatable, but more complicated because it requires to compute the
derivatives of a system of three implicit equations (see
\refapp{app:cep-calc}). The calculation of these derivatives will
become even more complicated (for example the dimensionality of the
system will increase) when more realistic model, with vector channel
interactions or with the Polyakov loop, will be used.

If we choose this particular model it is precisely because of its
simplicity hence besides some results relevant for the physics of the
CEP, this work is also a benchmark for further studies: in this model we
are able to cross check analytically large part of our numerical results.

For this reason we choose to compute the sensitivity using a
Monte-Carlo setup. It has the advantage that for higher dimensionality
of the CEP system (thus for more realistic model) it is well known a
Monte-Carlo will nicely scale whereas an analytic calculation would
become uninteresting (it is common that analytic derivatives formula
are not well suited for a computation because of differences or ratio
of almost identical terms) and a numerical calculation could become
too time consuming (it is the famous problem of equidistant sampling
versus random sampling in a high dimensionality space).

Besides, since we use the statistical error propagation formula, the
Monte Carlo is a natural framework to compute mean and standard
deviation. The last advantage of Monte Carlo at non vanishing
dispersion is the fact that one can visualize the data and have a
better understanding of the dispersion pattern of the CEP or the
correlations (as it will be discuss in the last section).

The details of the Monte-Carlo setup are only relevant in the case
of a finite dispersion of the inputs; they will be discuss there.

\subsection{Results on the sensitivity of some NJL predictions}

Let us come to the first results of this work.

\noindent
We choose the following value for the inputs:
\begin{equation} 
  \label{eq:mean-input-values}
  \begin{pmatrix}
    m_\pi \\
    f_\pi\\
    \qbarq^{1/3}
  \end{pmatrix}
  = 
  \begin{pmatrix}
    137 & \textrm{MeV} \\
    93 & \textrm{MeV} \\
    -315 & \textrm{MeV}
  \end{pmatrix}
\;.
\end{equation}
The value for the quark condensate is equivalent to choose the
$u$ condensate as \mbox{$\ubaru^{1/3} \simeq -250$ MeV}. Such value is
in agreement with limits extracted from sum rules, \\
\mbox{$190 \;\mathrm{MeV} \le - \langle \bar{u}u\rangle^{1/3} \le 260
  \;\mathrm{MeV}$} at a renormalization scale of $1$ GeV
\ccite{Dosch1998173}, and \mbox{$\langle \bar{u}u\rangle^{1/3} = -270
  \;\mathrm{MeV}$} at a renormalization scale of $2$ GeV
\ccite{BordesJHEP2010}. This value is also in agreement with recent
result of lattice calculations: \\
\mbox{$\langle \bar{u}u\rangle^{1/3} = -269(08)$ MeV}
\ccite{Aoki:2013ldr}.

\subsubsection{Parametrization}

In our context the true inputs of the model are the phenomenological
observables $m_\pi$, $f_\pi$ and $\qbarq$ from which the physical
parameters are uniquely defined. So the model parameters also have a
sensitivity.  As noticed by \ccite{Reinhard:2010,Fattoyev:2011}, a
variation around the solution of the inverse problem also contains
information on the accuracy of the
model. Very large sensitivity would mean that from the very beginning
the model cannot be used to do predictions because the whole model
itself (and not one of its predictions) is ill-conditioned (let us
recall that by model we mean the Lagrangian, its approximation
and also its inputs). We will see in the
section \ref{subsec:qq_var} a case where $\alpha \rightarrow \alpha_c$
($\alpha_c$ is the limit value of $f_\pi^3 / \qbarq$ where the inverse
problem cannot be solve anymore, see \refapp{app:inverse-pb}): all
sensitivities diverge at that point, even the parameters one. At this
specific point, the model in itself becomes ill-posed and nothing can
be done with it.
A correlation analysis would also be interesting: very low correlation
of the parameters with  the inputs would be a signal that the chosen inputs
are not the relevant  ones to constrain the parameters
\ccite{Fattoyev:2011}.

In \reftab{tab:sensitivity} one can read that we are rather safe with
the parametrization, the magnitude of sensitivities being below 5.  

It is worth noticing that a symmetry relation provide us a
very easy way to compute $\Sigma(m_0)$ but also to get more
information. An approximate value is easily obtained: at first order
in the bare mass $m_0$ the GMOR relation $m_\pi^2 f_\pi^2 = -m_0
\qbarq$ leads to $dm_0/m_0 = 2 dm_\pi/m_\pi + 2 df_\pi/f_\pi - 3 dc/c$
hence $\Sigma(m_0) = \sqrt{17} = 4.12$. By writing the differential we
see that the $m_0$ sensitivity is decomposed in 3 almost equal source
terms for the sensitivity (namely the 3 logarithmic derivatives). From
this decomposition one can learn that $m_0$ will be almost equally
sensitive to variations of any of the inputs. By the examination of
\refeq{eq:app-inverse-problem-sys-njl-1} it is not obvious at first
glance that this should be the case. One could think that $m_0$ is
strongly sensitive to $m_\pi$ but due to the GMOR relation it is not
the case. Then, even at the simplest level, we think that the study of
sensitivities with respect to the inputs and their sources (related to
the partial derivative) can bring information that may be difficult to
find based on purely physical argument. For the case of $m_0$ the GMOR
relation is enough to reveal the hidden link between observables but
we will see that the more complicated the prediction, the less obvious
this kind of link can be found without a sensitivity calculation.

Finally let us stress again the first strong strength of a sensitivity
analysis: if the sensitivities of the parameters are large it is
meaningless to even try to use the model.


\begin{table}[!ht]
  \centering
  \begin{tabular}{c|c|l|l|}
    \cline{2-4}
    \multicolumn{1}{l|}{} &
    \multicolumn{2}{c|}{\begin{tabular}[c]{@{}c@{}}Sensitivities\end{tabular}}
    & Values \\ \hline
    \multicolumn{1}{|c|}{\multirow{3}{*}{Parameters}} & $\Lambda$ &
    $2.83$ & $0.653$ (GeV) \\ \cline{2-4} 
    \multicolumn{1}{|c|}{} & $m_0$ & $4.11$ & $0.0051$ (GeV)\\ \cline{2-4} 
    \multicolumn{1}{|c|}{} & $G\Lambda^2$ & $3.32$ & $2.11$\\ \hline
    \multicolumn{1}{|c|}{\multirow{3}{*}{\begin{tabular}[c]{@{}c@{}}In-vacuum\\
          predictions\end{tabular}}} & $m$ & $6.72$ & $0.313$ (GeV) \\ \cline{2-4} 
    \multicolumn{1}{|c|}{} & $m_\sigma$ & $6.41$ & $0.642$ (GeV) \\ \cline{2-4} 
    \multicolumn{1}{|c|}{} & $g_{\pi \bar q q}$ & $5.97$ & $3.37$\\ \hline
    \multicolumn{1}{|c|}{\multirow{2}{*}{\begin{tabular}[c]{@{}c@{}}In-medium\\
          predictions\end{tabular}}} & $\Tcep$ & $71.5$ & $0.0299$ (GeV) \\ \cline{2-4} 
    \multicolumn{1}{|c|}{} & $\mucep$ &  $1.05$ & $0.327$ (GeV)\\ \hline
  \end{tabular}
  \caption{ \label{tab:sensitivity} {\small Sensitivities of the
      parameters, in-vacuum predictions and in-medium predictions
      considering infinitesimal changes of the inputs. The
      sensitivities of the parameters, of the in-vacuum predictions
      and of $\mucep$ are close to $1$. The sensitivity of the
      temperature coordinate of the CEP is very large.
      These values were computed numerically with a Monte-Carlo.
  }}
\end{table}

\subsubsection{Prediction sensitivities}

Considering that our model is constrained in vacuum, in
\reftab{tab:sensitivity} we organize the sensitivities (top to bottom)
from the expected lowest values (for the parameters) to the expected
greatest values (for the in-medium predictions). \\

We began our discussion on the usefulness of the sensitivity analysis
by the sigma meson mass and the pion-quark-antiquark effective coupling
constant in vacuum (see \refeq{eq:vac-predictions}).

As expected, the sensitivities of the in-vacuum predictions are larger
than the one of the parameters, but are still not too large
($<7$). The phenomenological conclusions that can be done within this
model in the in-vacuum mesonic sector are rather safe.

Let us take a closer look on the sigma mass: In
\refapp{app:msigma-sensitivity} we compute analytically $\Sigma(m_\sigma)$
(hence checking that the present Monte Carlo setup is correct) and
obtain the result for the differential $ \dd m_\sigma = 0.21 \dd m_\pi + 34
\dd f_\pi - 8.2 \dd c$. \\
First of all we see that the sensitivity coming from the pion mass
partial derivative is negligible and this is not surprising: it is
well known that the Goldstone theorem is approximately realized in NJL
due to the the small values of the bare quark mass so the sigma mass
is essentially twice the dressed quark mass. In fact, for this reason
all observables are barely sensitive to the pion mass (except
of course if they correlate directly with $m_0$). \\
But the sensitivity analysis can show relations that are more
difficult to predict based on simple symmetry arguments.  For example,
since $m_\sigma^2 = 4m^2 + m_\pi^2$ and since $m$ is essentially due
to the chiral symmetry breaking in the NJL model by a quick
examination of the equation one could have expected that the sigma is
mainly sensitive to the condensate variation. The coefficients of this
differential shows that it is also sensitive to $f_\pi$ a fact that is
not easily read in the NJL model due to the non-linear relation
between the scale $\Lambda$, $f_\pi$ and the condensate. Let us stress
again that this simple model is used as an illustration of the
usefulness of this technique; the link between observables may not be
apparent by examination of the NJL equations but for the specific case
of $f_\pi$ one can see for example in a linear sigma model that the
PCAC imposes an approximate proportionality between the vacuum
expectation value of the sigma field and $f_\pi$: $v^2 = f_\pi^2 (1 +
o(m_\pi^2) )$ hence a strong correlation between $m$ and $f_\pi$. It
is another example of the fact that sensitivity or correlation
analysis can bring to light such
relations, hidden by the non linearity of the equations. For the CEP, such
correlations become very difficult to predict a priori (as we will see
in the following) hence
the analysis is a useful one for phenomenology. \\

Finally let us conclude by examining the sensitivities of the
in-medium predictions of $\Tcep$ and $\mucep$. They are both
surprising. On the one hand the sensitivity of the $T$-coordinate of
the CEP is very large ($\sim 70$) and the one of the $\mu$-coordinate
is very close to $1$, even more close than the
sensitivities of the model parameters. \\
Concerning the $\Tcep$, the conclusion is quite easy. The model, that
consists of the approximation schemes and of the phenomenological
inputs, is ill-conditioned for the prediction of the temperature
coordinate of the chiral CEP: we try to push too far the
model from the vacuum where its parameters are constrained. This means
that no consistent conclusions can be drawn concerning $\Tcep$ in this
context (we will detail this in the section devoted to small but
finite variation of inputs). \\
The conclusion is completely different for $\mucep$. Its close to
unity sensitivity means that the physics that is implemented in this
simple NJL model, \idest chiral symmetry and scalar sector, is part of
the physics that constrains the chemical potential coordinate of the
CEP. This does not mean that the $\mucep$ in nature is the one that is
predicted by this NJL model but this means that the chiral physics
generated by the scalar channel seems to be very relevant to this
prediction.  For example adding new physical contents to the model,
such as vector channel interactions, will certainly change the
position of $\mucep$ in the phase diagram (essentially it will shift
it \ccite{NJLrev_klevansky}) but we have reason to believe it will not
alter drastically the sensitivity of $\mucep$ and not solve the
sensitivity problem of $\Tcep$\footnote{This conjecture is based on
  \ccite{Lourenco2012} Fig. 1 where the CEP seems to remain not very
  well constrained.}.  To solve this problem
one probably need in-medium constraints as we will check in a future
work.

\subsection{Sensitivities for different values of the quark
  condensate \label{subsec:qq_var}}

Since the value of the quark condensate is less well known when
compared to the pion mass and the pion decay constant, we plot the
sensitivities of $m$, $\Tcep$ and $\mucep$ as a function of the
value of the condensate with $m_\pi$ and $f_\pi$ fixed to their
values given in \refeq{eq:mean-input-values}.

\begin{figure}[!ht]  
  \begin{center}
    \includegraphics[width=0.4\textwidth]{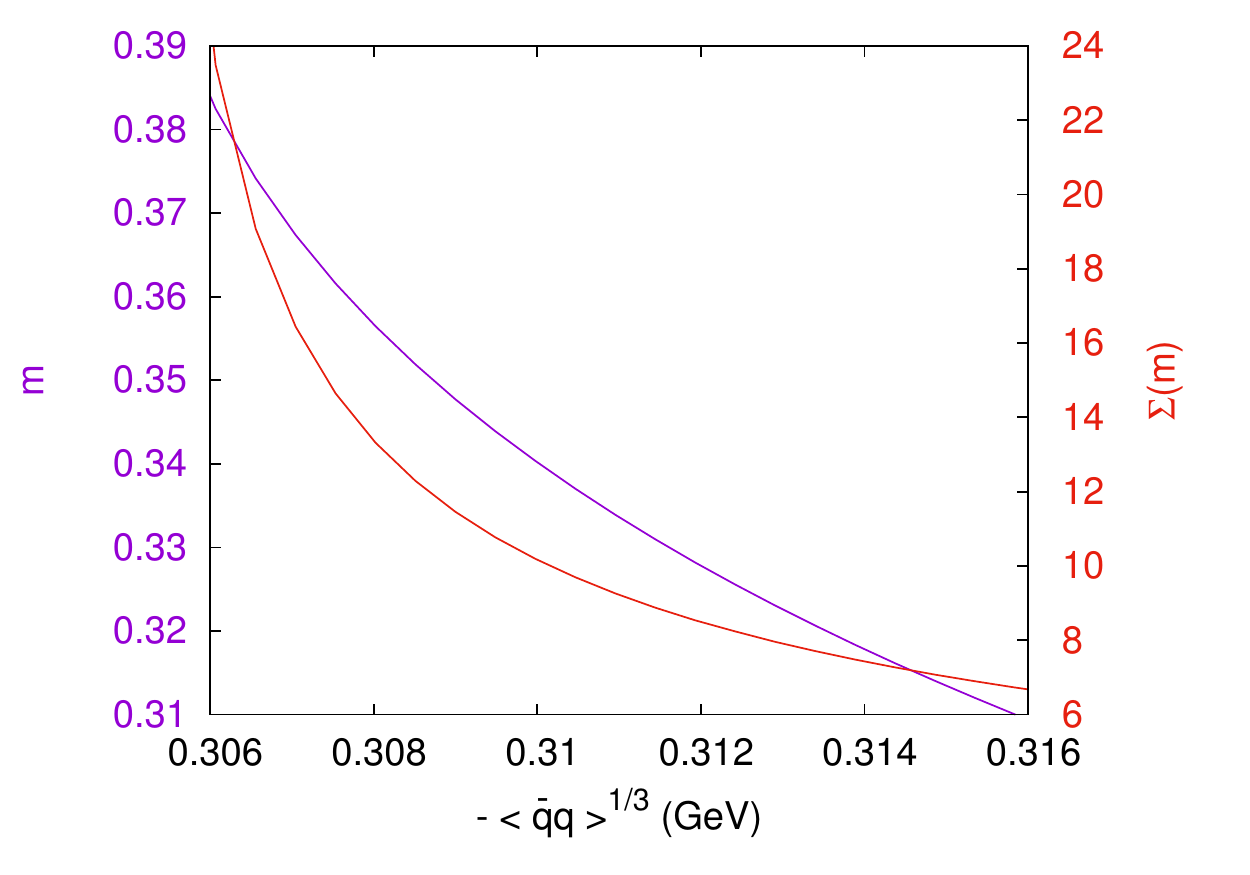}\\
    \includegraphics[width=0.4\textwidth]{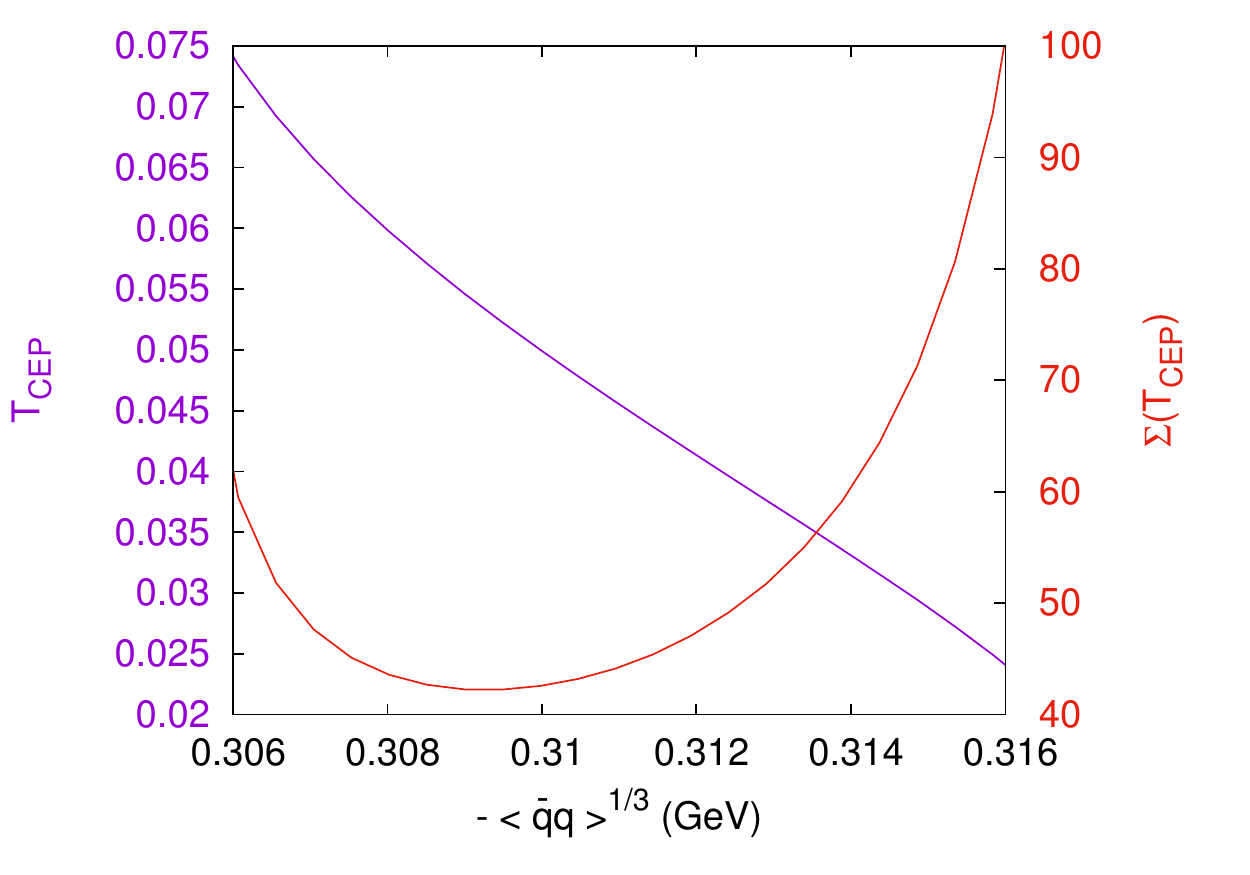}\\
    \includegraphics[width=0.4\textwidth]{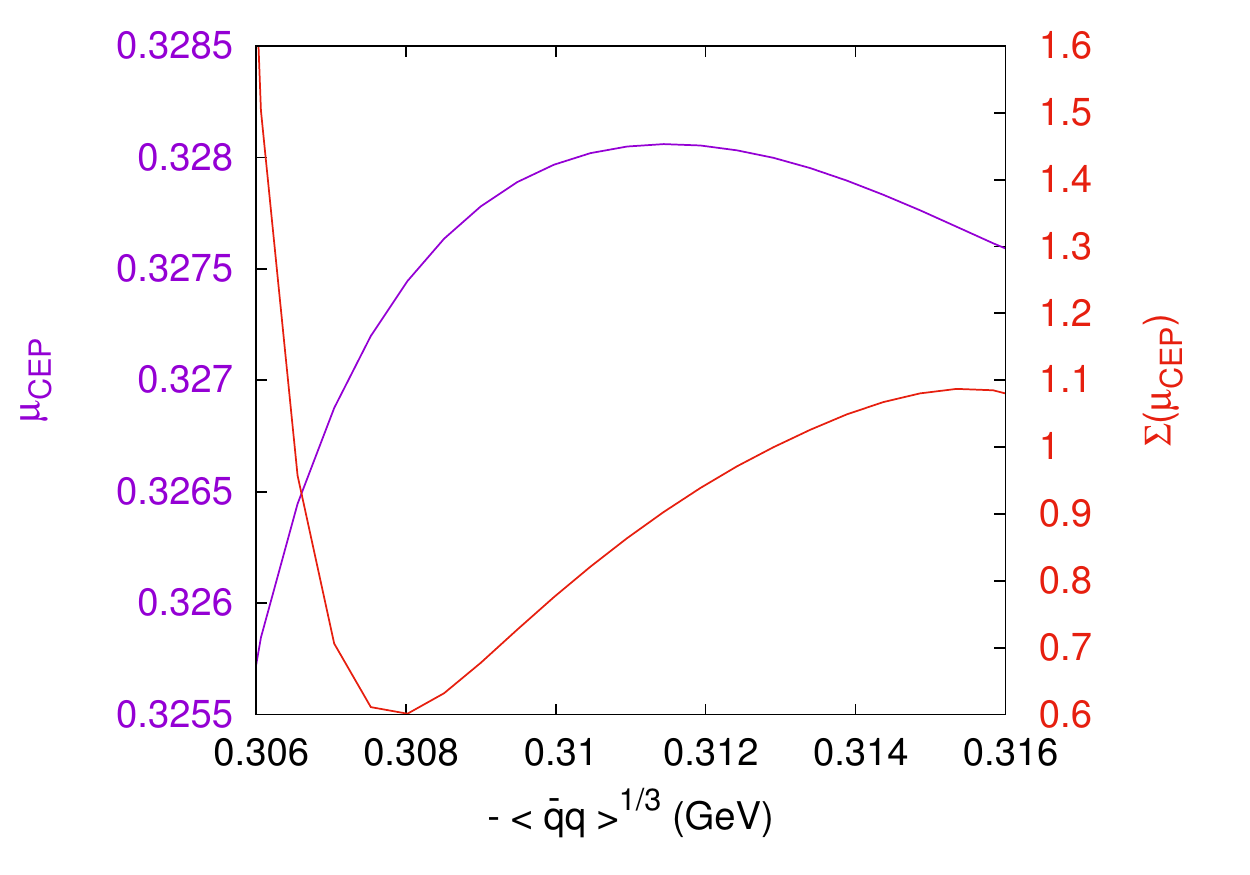}
  \end{center}
  \caption{\label{fig:sensitivities-fixed-qq} {\small The constituent
      quark mass $m$ (top), the temperature coordinate of the CEP
      (middle) and the chemical potential coordinate of the CEP
      (bottom) as well as their corresponding sensitivities are
      plotted as functions of $- \qbarq^{1/3}$. We limit the range
      of the quark condensate such that the inverse problem always has
      a solution and that the CEP always exists. When the quark
      condensate varies in a $10$ MeV range, the constituent quark
      mass varies in a $85$ MeV range, $\Tcep$ varies in a $60$ MeV
      range and $\mucep$ only in a $2.5$ MeV range. These windows are
      closely related to the values of the sensitivities that are
      average for $m$, large for $\Tcep$ and close to $1$ for $\mucep$}}
\end{figure}

In \reffig{fig:sensitivities-fixed-qq}, the constituent quark mass,
the temperature coordinate of the CEP, its chemical potential
coordinate and their corresponding sensitivities are plotted as a function of 
the value of the quark condensate in the range $[306 , 316]$ MeV. At
$c = 306$ MeV, $\alpha \simeq \alpha_c$ and the inverse problem does not have
a solution anymore; at $c \simeq 316$ MeV, $T \simeq 0$: the CEP
disappears from the phase diagram. As we mentioned earlier, we see
that the sensitivities diverge at the lower limit of the range. It is
the case for all quantities: in fact at this point, the model is
ill-posed and the parameters cannot even be fixed.

On the lower panel of \reffig{fig:sensitivities-fixed-qq}, one can see
that $\mucep$ only varies in a $2.5$ MeV range and this is coherent
with what we expect from the behavior of a quantity with such a small
sensitivity: the prediction of $\mucep$ is very stable even with
respect to large variation of the condensate (this is also confirmed
by our calculation of the correlation of $\mucep$ with the condensate
that turn out to be small as can be seen in \reftab{tab:correl} ;
correlations will be discussed in the last part). On
the contrary the constituent mass varies in a range of $85$ MeV (upper
panel) and $\Tcep$ in a $60$ MeV range (middle panel). This is
coherent with the fact that $m$ and $\Tcep$ have larger
sensitivities. As a side note, we see that $\Sigma(T)$ is large on the
whole interval signaling a fundamental problem for this model
in order to constrain the temperature and give an
accurate prediction; also it even diverges just before the CEP disappears. \\
On this range $\Delta m / m \simeq \pm 11\%$ and $\Delta T/T \simeq \pm 100
\%$.  Of course these values are not exactly the sensitivities
previously computed. Sensitivities are local quantities (related to a
gradient with respect to the 3 directions in the inputs space) but it
shows that there is a correct agreement between this finite variation
and an extrapolation based on the first order of a Taylor expansion
(whose coefficient is related to the sensitivity). It shows that the
problem is sufficiently linear (at least in the $c$ direction) around
the input values for the sensitivity to be a useful quantities even in
this non linear problem. We will use this fact when studying finite
variations of all 3 parameters.

Since the quark condensate is closely related to the dynamical
generation of the mass and the latter being a relevant phenomenon for
the creation of a first order phase transition at zero temperature we could
anticipate that $\mucep$ and $m$ would varies accordingly when
$\qbarq$ only is varied. By comparing the upper and lower panels,
one can see it is not the case: the larger sensitivity of the
constituent mass leads to a large variation of the mass but the
chemical potential is remarkably stable. This result shows how the
non-linearity of the inverse problem may affect the outcome in a non
trivial way that is revealed by the sensitivity. We will see that the
chemical potential is strongly correlated with the pion decay constant.

From top of \reffig{fig:sensitivities-fixed-qq}, one notices that, at
fixed $f_\pi$ and $m_\pi$, the constituent quark mass reduces when $c$
raises. From $m \simeq 2 G c^3$, one could expect that $m$ would
increase when $c$ increases, and so should $\mucep$ since it would
take more density to destroy the condensate. Here, the calculation
results in a decrease of $m$ while $\mucep$ is non-monotonic. This
is another interest of varying the inputs of the model and not the
model parameters: the phenomenology may vary counter intuitively when
one realizes that the true inputs are the datae. In fact, when $m_\pi$
and $f_\pi$ are fixed, $G \equiv G(\qbarq)$, and then $m \simeq
-2G(\qbarq) \qbarq$, and the non linear behavior of $G$ makes wrong
the first conclusion.

Finally, let us recall that this analysis is rooted in the nuclear
physics community (both theory and experiments).  It can serve as a
guide for experiments. As noticed by \ccite{Fattoyev:2011} the
correlation analysis can almost systematically determines which
observables are the best to provide constraints on parameters and then
experiments can concentrate on the most readily accessible of these
observables. In our case and as a though experiment, let us suppose
one moment that the scalar channel is the only relevant one, that the
CEP has been shown to exist (for example if it is shown that at zero
temperature in compact star phenomenology the chiral transition is
first order) and the pion properties are well known but the chiral
condensate has not been measured. The result of this section shows
that the chiral condensate must be searched in the range $[306 , 316]$
MeV even if we cannot point where the CEP should be in the phase
diagram due to the temperature sensitivity.

\subsubsection{On the choice of the dispersion pattern}

We have chosen to take equal relative dispersion of the inputs.  Of
course choosing equal absolute dispersion changes the results.  With
the values we use for the inputs it means that the condensate varies
approximately 3 times more that the other inputs. It can be an
informed choice, for example if one estimates that since it is less
known it should vary more. In any way to check if our conclusion where
change by this choice we also computed with absolute dispersion.  None
of the above conclusions are changed by the other choice.

%
%


\section{Consequences of small but finite deviations of the inputs}

\subsection{Why finite variations are relevant}

In our case of an exact inverse problem a large value of $\Sigma(T)$, that
is a calculation close to a ill-posed one, may be harmless since it is
computed at vanishing dispersion and the parameters fixing procedure
is exact. Indeed, if the inputs are very accurately computed in the
model and very well measured, the outcome may vary only slightly (the
situation would be worse if we used a $\chi^2 \neq 0$ to parametrize the
model since there would be no way to fix the parameter to reproduce
the exact inputs). However, as an illustration of our previous analysis
we will show that for our calculation of the CEP it is
unacceptable. We will see that very small variations (0.6\%) around
the vacuum will already completely change the physics of the model,
namely with such small variations the CEP may or may not exists. The
problem is that the inputs are neither very well measured nor
accurately computed. 

On the one hand the quark condensate value is not very well known as
explained earlier. 
On the other hand let us stress that when we compute the sensitivity
in this model we mean the NJL Lagrangian and also its
approximations. By comparing our results with results obtained with a
less approximate treatment we can evaluate roughly the order of
magnitude of the systematic errors generated by the approximation.

For example, relaxing the quasi-Goldstone boson approximation (we
reinstate $k^2 = m_\pi^2$ in \refeq{eq:mpi}) we find $m_\pi = 135.6$
MeV accounting for a variation of about 1\%.

Let us consider the next order in a $1/N_c$ expansion as the
meson-loop approximation (MLA) \ccite{Oertel2000}.  In this work, the
correction on the pion properties were found around 5\% (the value
depending on details of the model calculation).  In the framework of
the inverse problem, it means a re-parametrization has to be made to
get back the correct vacuum phenomenology. Here we do the assumption
that, during the re-parametrization procedure, the previously computed
mean field sensitivity may already generate a variation for a
prediction $X$ of about $\Sigma(X) \times 5\%$.

We believe that our calculation of the sensitivity and a rough
approximation of the contribution of the next order, for the inputs,
is able to determine if this correction will be likely to damage the
current calculation.


\subsection{CEP unpredictability}

In order to see the concrete effects of large or small sensitivities
of the predictions, we allow the phenomenological inputs to vary in a
small range given by a relative dispersion $p = 1\%$ of the mean
values given in \refeq{eq:mean-input-values}. As we have seen this
value is rather conservative considering the MLA estimation. It is
worth writing explicitly the range where our value will fluctuate. The
range is rather small (especially when looking to various NJL model
parametrization in the literature) and yet the physics will be
completely changed:
\begin{equation}
  \begin{matrix}
    m_\pi \; &\in& \; &[&135.6\,&,&\, 138.4&]&  \\
    f_\pi \; &\in &\; &[& 92.07 \,&,&\, 93.93&]&\\
    \qbarq^{1/3} \; &\in& \; &[&-318.1 \,&,&\, -311.8&]& 
  \end{matrix}
\end{equation}

We will then see if the value of the sensitivities has a consequence
on the prediction with finite dispersion. Also additional information
can be acquired. The shape of the distribution, for example of the CEP
in the $(T - \mu)$ plane, can now be visualized, together with
correlation plots. \\

We must now explain precisely our Monte Carlo setup (the previous
calculation were done in the exact same way with $p = 0.005\%$
and we checked that it was small enough for the calculations
to extrapolate toward $p=0$). \\
A set of $n$ input points is generated following a given probability
density (the choice of the density is irrelevant at vanishing
dispersion).

The uniform distribution $\rho^u(X)$, which is suited for analysis of
deterministic errors, is a constant around its mean value $\bar{X}$:
\begin{equation}
  \label{eq:uniform-distribution}
  \rho^u(X) = {\cal N} \theta(X-X_{\mrm{max}}) \theta(X_{\mrm{min}} - X) \;,
\end{equation}
with $\theta$ the Heaviside function, $\cal N$ a constant that normalizes
the density to one and where $X_{\mrm{min}} = (1-p) \bar X$, 
$X_{\mrm{max}} = (1+p) \bar X$, where $p=1\%$. \\
The Gaussian distribution $\rho^G(X)$ is usually used when supposing a
random variable normally distributed with a standard deviation
$\sigma$.  The interest of the Gaussian distribution is that its wings
will allow us to explore points that are not in the uniform
distribution. To compare the results obtained with the uniform
distribution, we used $\sigma = p \bar X$ and then:
\begin{equation}
  \label{eq:gaussian-distribution}
\rho^G(X) = {\cal N} e^{ (X-\bar X)/2\sigma^2 } \;,
\end{equation}
where again $\cal N$ is the normalization of the distribution. Using
these definitions we checked that the results do not change
qualitatively when using the uniform or the Gaussian distribution.

Then, the inverse problem is solved leading to $n$ sets of
parameters. For each of these sets of parameters the in-vacuum as well
as the in-medium predictions are computed. At the end, distributions
for $m_\sigma$, $g_{\pi \bar q q}$ and also for $\Tcep$ and $\mucep$
are obtained. For each of the distribution $\rho_X$ of the prediction
$X$ a mean value $\bar X$ and a standard deviation $\sigma(X)$ can be
computed. The sensitivity of the prediction $X$ \refeq{eq:sensitivity}
becomes:
\begin{equation}
  \label{eq:mc-sensitivity}
  \Sigma(X) = \frac{\sigma(X)}{\bar X} \frac{1}{\sigma_{rel}^{I}} \;.
\end{equation}

\subsubsection{Distributions of the model parameters}
\label{ssec:Distributions of the model-parameters}

\begin{figure}[t]
\includegraphics[width=0.4\textwidth]{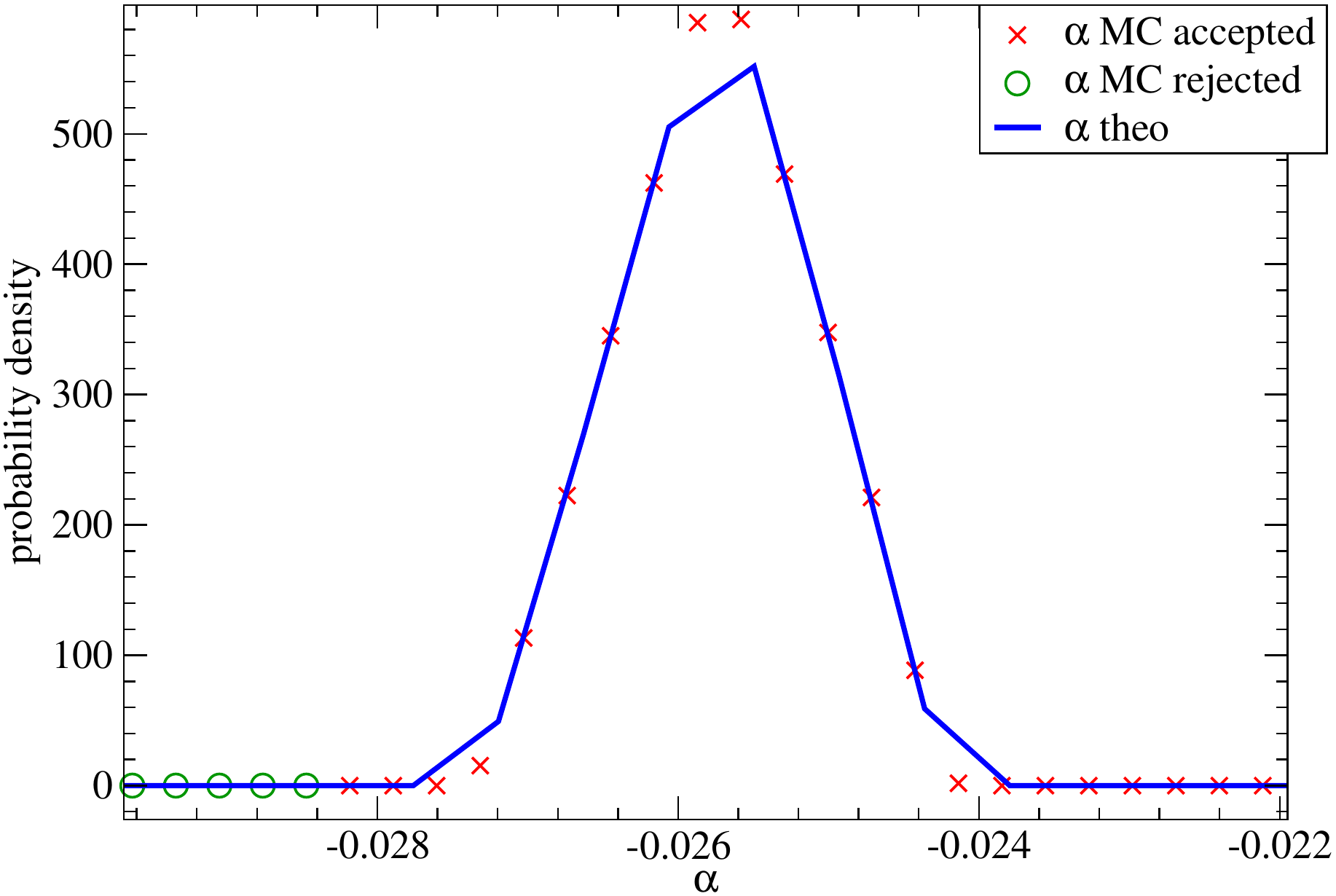}
\caption{ \label{fig:alpha-dist} Distributions of $\alpha$ with a
  dispersion $p = 1\%$ of the inputs for the uniform distributions.
  The full line represents the theoretical density given by
  \refeq{eq:app-rho-alpha-theo}, and the circle and cross are the
  value of a normalized histogram (the integral is one)
  obtained with the Monte Carlo
  simulation ($n=50^3$). The crosses are the accepted parameters
  (satisfying $\alpha > \alpha_c$), and the circles are the rejected
  parameters ($\alpha < \alpha_c$).  }
\end{figure}

The probability distribution of $\alpha$ is plotted in
\reffig{fig:alpha-dist} where the theoretical distribution, that is
calculated in \refapp{app:ana-prob-dist}, is also shown as a cross
check.

\begin{figure}[!h]
\includegraphics[width=0.4\textwidth]{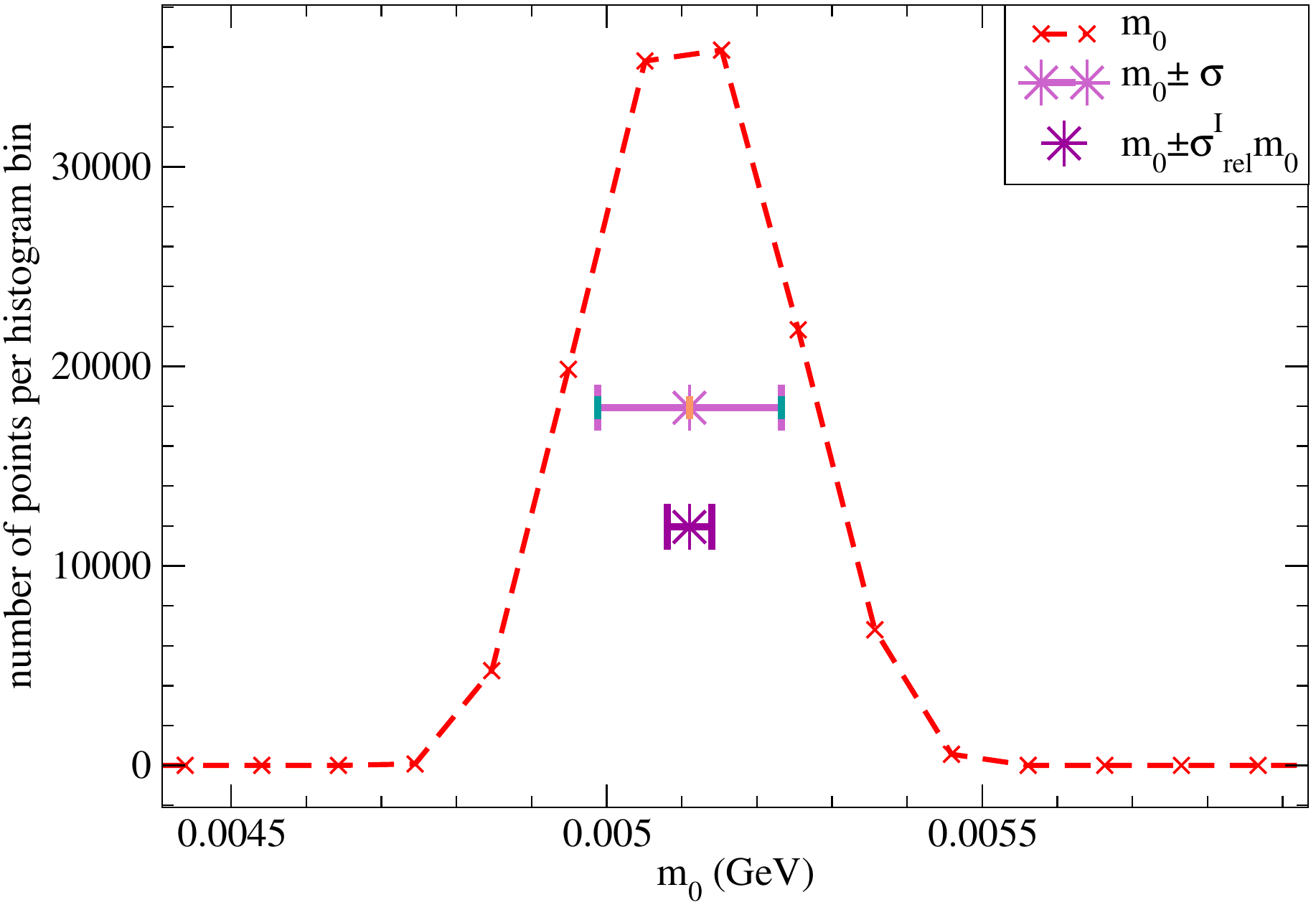}
\includegraphics[width=0.4\textwidth]{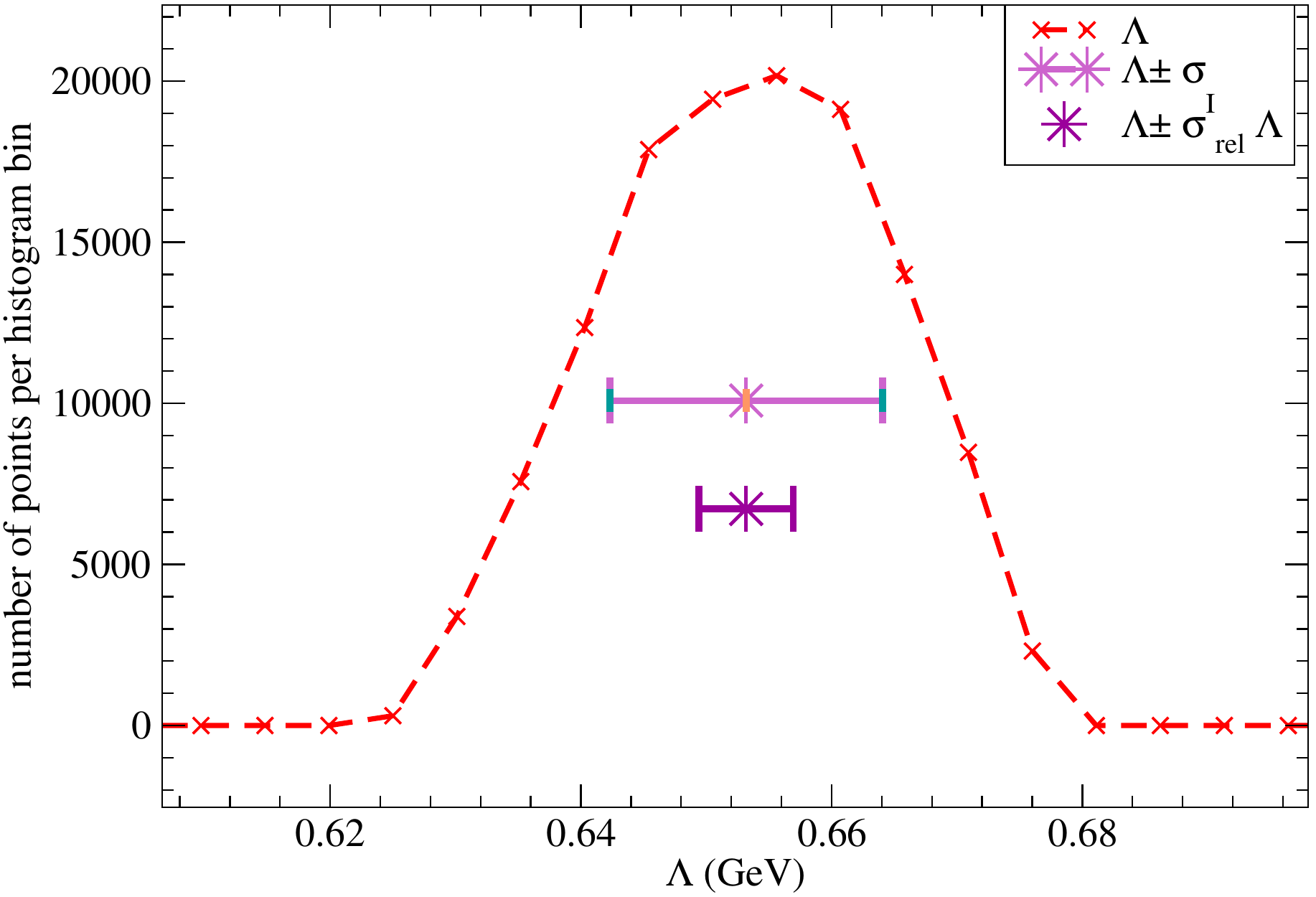}
\includegraphics[width=0.4\textwidth]{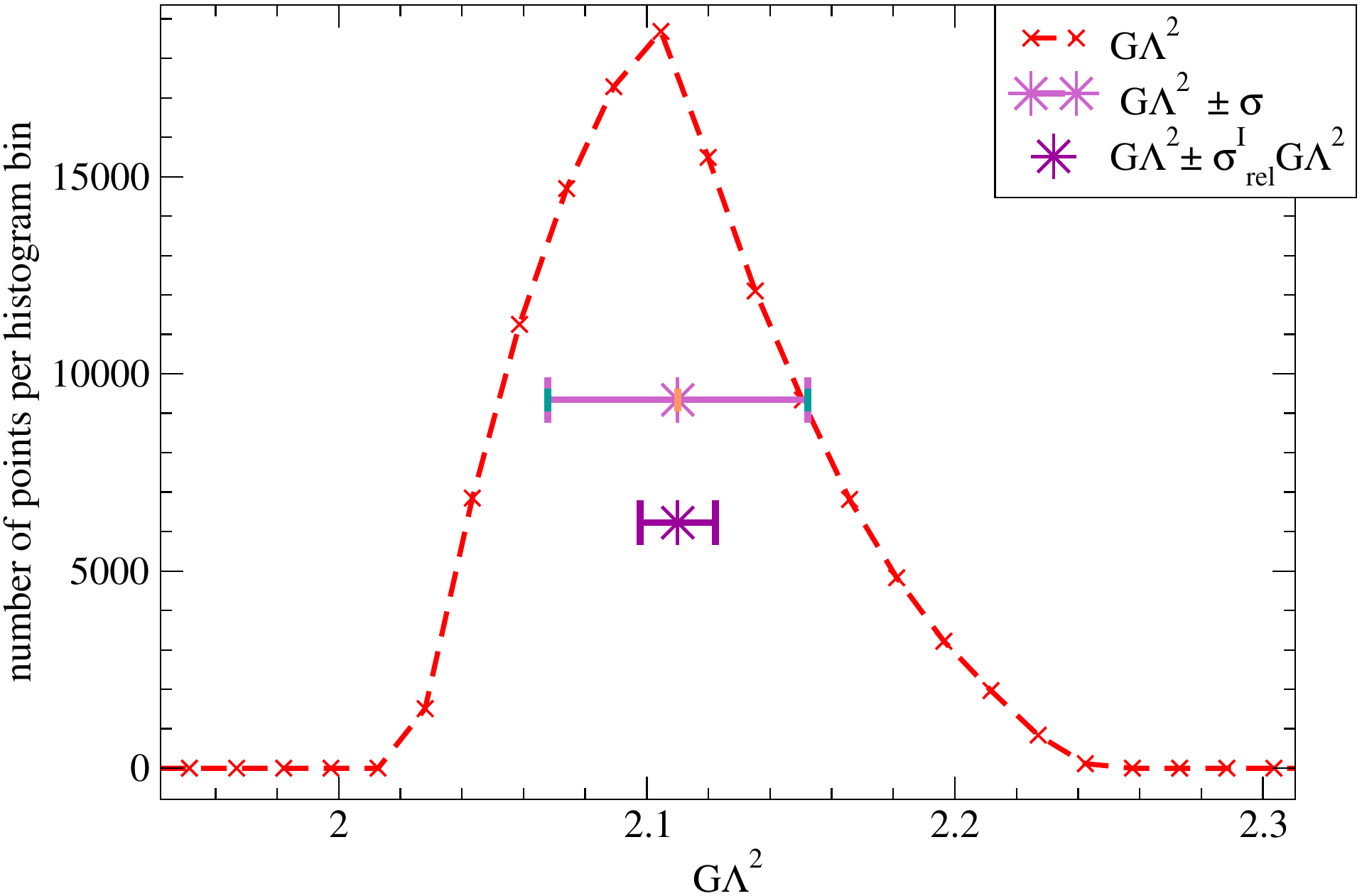}

\caption{ \label{fig:params-dist} The crossed lines represent the
  distributions of the dimensional parameters of the NJL model
  (un-normalized histogram whose integral is the number of points):
  $m_0$ (top), $\Lambda$ (middle) together with $G \Lambda^2$ (bottom).
  The results are obtained
  with the Monte Carlo simulation with a variation of $1\%$ of the
  inputs for the uniform distribution and $n=50^3$.  The top star
  point and the error bar represent the mean and standard
  deviation. The bottom star point has error bar whose value is the
  mean times the standard deviation of the inputs, $S(X)=\overline
  X~\sigma_{rel}^I$ , a way to visualize the sensitivity of the output
  with respect to the dispersion of the inputs (see discussion in
  text). Notice how these error bars are always smaller (especially
  for $m_0$) than the standard deviation illustrating how an initial
  dispersion of the inputs translates in larger deviation of the
  outputs.  }
\end{figure}

In \reffig{fig:params-dist} we plot the obtained distributions of the
NJL model parameters. We also represent the mean and standard
deviation of the parameters and check that the statistics was large
enough to have well controlled errors. For each random variable $X$,
another point is represented on the figures whose error bar are
$S(X)~=~\overline X~\sigma_{rel}^I$.  This point allow us to compare the
relative standard deviation of a quantity to $\sigma_{rel}^I$. When
$\sigma(X) > S(X)$ we have a visual estimation that the chosen
dispersion for the inputs results in a larger dispersion for this
output. This point is related with the sensitivity, indeed $\Sigma(X) =
\sigma(X) / S(X)$.

The bare quark mass value is between $4$ and $6$ MeV, typical values
found in literature. It is worth noticing that with a sensitivity of
4, if one would like to tackle the difficult problem of the evaluation
of the bare quark mass with this model and measurements of pion
properties and the condensate, the value would be affected by a large
uncertainty.

For what concerns $G$ there is a small absolute dispersion of the
parameter and one notices a very sharp low cut of its value.  The
sharp cut induces a strong asymmetry of the density as can be seen
from the position of the mean.  The values of $G\Lambda^2$ are located
around $2.1$ with $\sigma(G) \simeq 0.1$.

Finally, $\Lambda$ displays a quite large standard deviation and its
typical values are between $600$ and $700$ MeV. The latter value is a
bit large compared to the usual parametrization of the model.

\subsubsection{Distributions of the sigma mass}
\label{ssec:Distributions of the in-vacuum predictions}

\begin{figure}[!h]
  \includegraphics[width=0.45\textwidth]{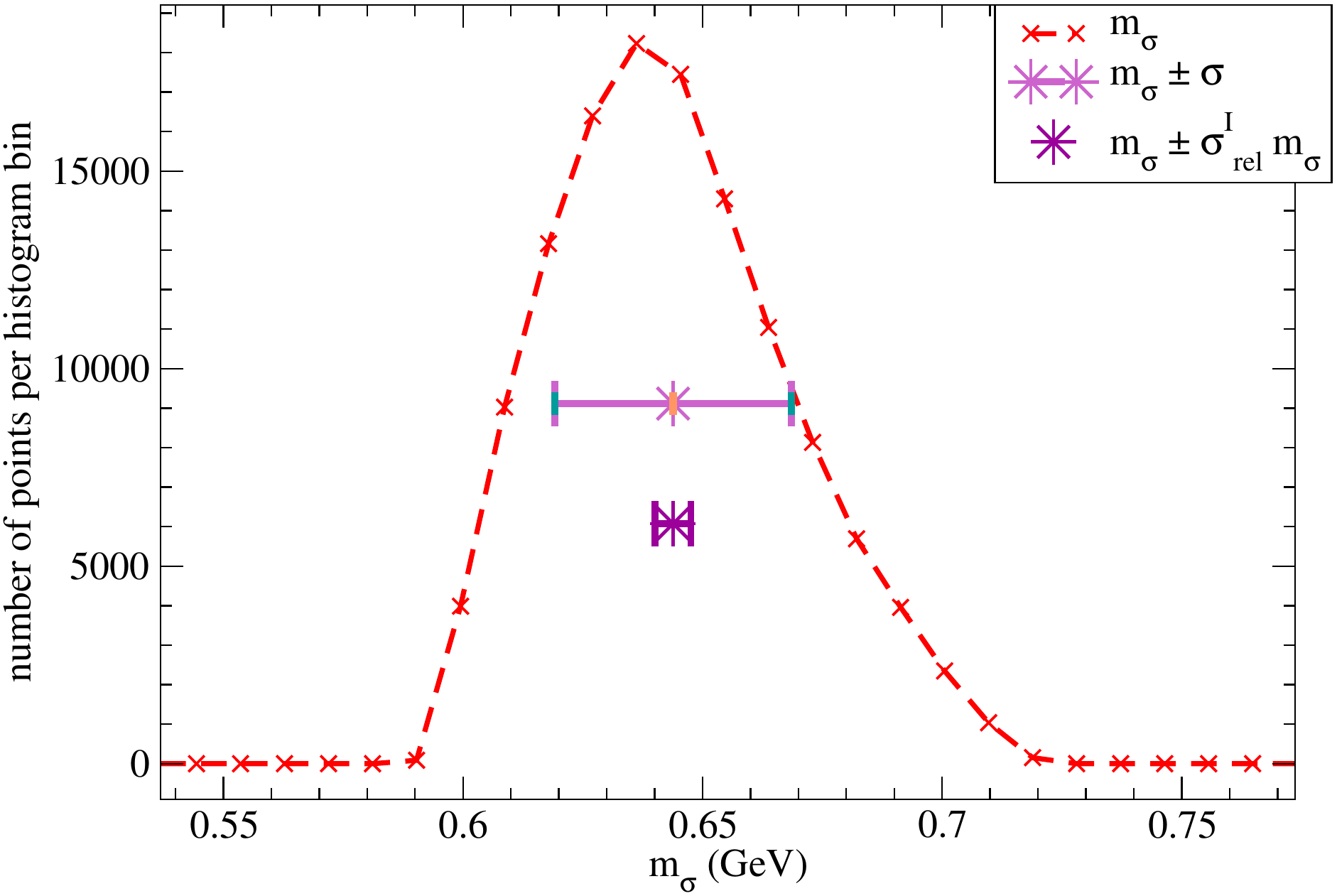}
  \caption{ \label{fig:sigma_mass} The crossed lines represent the
    distributions of $m_\sigma$
    (un-normalized histogram whose integral is the number of points)
    obtained with the Monte-Carlo simulation with variations of $1\%$
    of the inputs for the uniform distributions and $n=50^3$. The top
    stars and its error bars represent a visualization
    of the sensitivity (see \reffig{fig:params-dist} and text).
  }
\end{figure}

\reffig{fig:sigma_mass} presents the distribution of $m_\sigma$ and its mean
value and associated standard deviation. These quantities are also
gathered in \reftab{tab:in-vacuum-qual-results}, with the
corresponding values for the other prediction $g_{\pi \bar q q}$.  We
can notice that the shape of the density, even with the sharp cut of
the uniform parameter distribution, present a long tail for higher
value of the mass, not completely excluding value as high as 800 MeV
(the same tail can be seen for $G\Lambda^2$). As we can see from the table,
the results for $m_{\sigma}$ and for $g_{\pi{\bar{q}q}}$ are in accordance
with the phenomenology of simple quark models ($m_\sigma \simeq 600$ MeV
is the expected value for the sigma mass in this framework; it cannot be
compare to the experimental scalar meson as discussed for example in
\ccite{Celenza2000,Celenza2001}). Furthermore, the dispersions of these quantities
are reasonable (less than $7\%$).



\begin{table}[h]
  \centering
  \begin{tabular}{|l|r|}
    \hline
    $ \overline{m_\sigma}$   & $(0.6439 \pm 0.0003)$ (GeV) \\ \hline
    $ \sigma (m_\sigma) $     & $(0.0246 \pm 0.0002)$ (GeV) \\ \hline
    $\sigma(m_\sigma)/\overline{m_\sigma}$ & $3.82$ $(\%)$ \\ \hline\hline
    $ \overline{g_{\pi \bar q q}}$  & $ 3.3822 \pm 0.0001 $\\ \hline
    $ \sigma(g_{\pi \bar q q})$     & $0.209\pm 0.002$   \\ \hline
    $ \sigma(g_{\pi \bar q q})/ \overline{g_{\pi \bar q q}}$ & $ 6.18$
    $(\%)$\\ \hline
  \end{tabular}
  \caption{ \label{tab:in-vacuum-qual-results} 
    Results obtained for the in-vacuum predictions $m_\sigma$ and
    $g_{\pi\bar q q}$, for the uniform distribution,
    with $n^3 = 50^3$ and $p = 1\%$.
  }
\end{table}

\subsubsection{Distribution of the chiral critical end point
  prediction}
\label{ssec:Distribution of the chiral critical end point prediction}

Let us come to the most striking result of this section. As explained
before, it may happen that the inverse problem has a solution but the
corresponding parameter set does not lead to a CEP.  With a dispersion
as low as 0.6\%, the CEP starts to disappear. Hence the sensitivity of
the CEP temperature is so large that the prediction is already spoiled
if one assumes only 0.6\% variations of the inputs. The existence of
the CEP (even if it exists when using the mean value of the inputs)
cannot be considered as a true prediction of this particular model;
the physical outcome of the model is completely changed. 

As a reference the number of obtained sets of parameter and the number
of calculated CEP for a dispersion of $1\%$ and $0.5\%$ is listed in
\reftab{tab:nb-CEP}.

\begin{table}[!ht]
\centering
\begin{tabular}{|c|c|c|c|}
\hline
\multicolumn{4}{|c|}{Uniform distribution} \\ \hline
$p$ (\%) & $n_{\textrm{sets}}$ & $n_{\textrm{CEP}}$ & $n_{\textrm{CEP}} / n_{\textrm{sets}}$  (\%) \\ \hline
1 & 3375 & 3066 & 91 \\ \hline
0.5 & 3375 & 3375 &  100 \\ \hline
\end{tabular}
\caption{
  \label{tab:nb-CEP}
    From the $n^3 = 15^3$ input sets generated with dispersion $p$ in
    the Monte-Carlo, $n_{\mathrm{sets}}$ parameter sets could be
    calculated (the solution of the inverse problem exists) and
    $n_{\mathrm{CEP}}$ admit a CEP.}
\end{table}

\begin{figure}[!h]
  \includegraphics[width=0.4\textwidth]{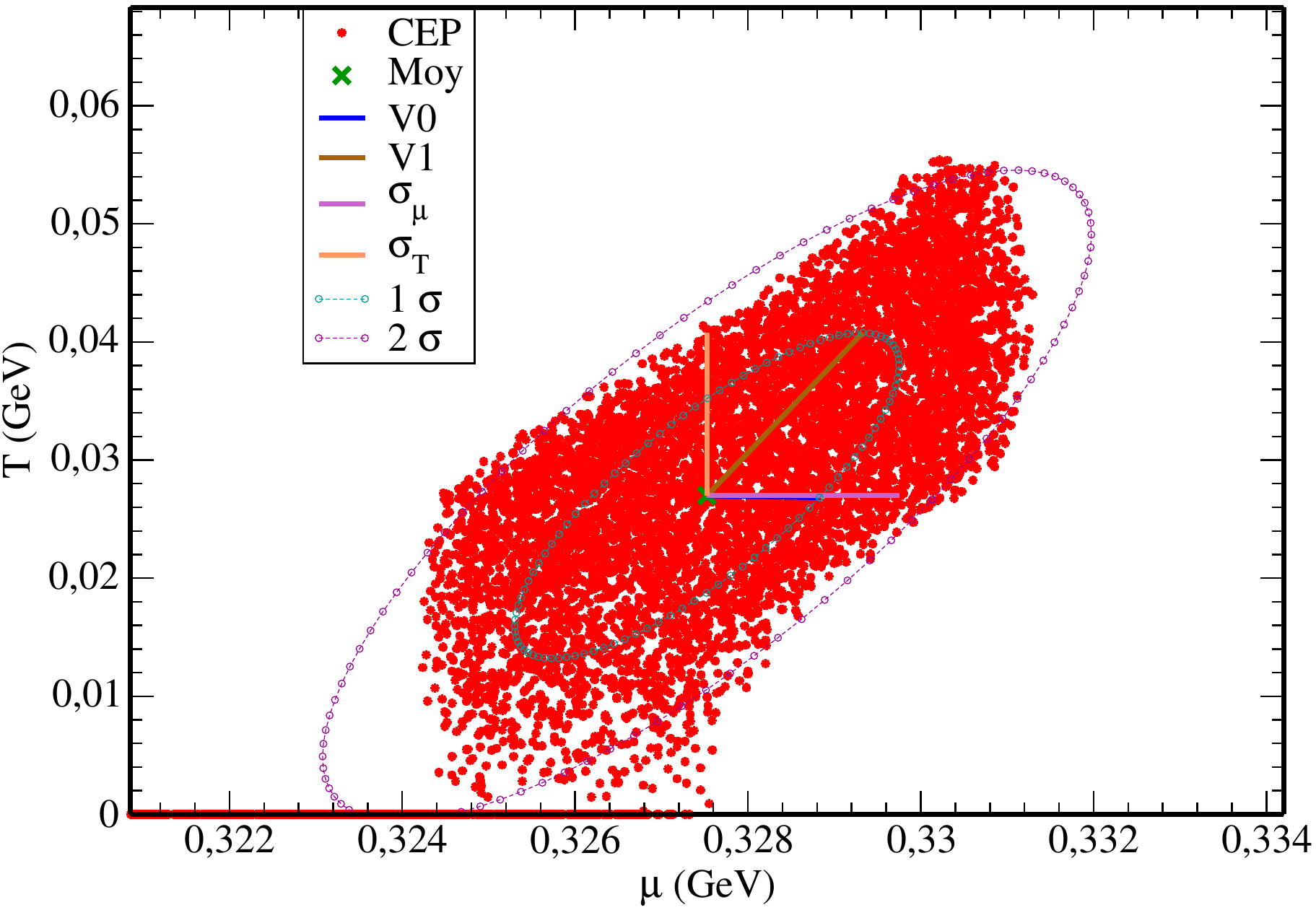}
  \caption{\label{fig:CEP-dist} Scatter plot of the CEPs obtained for
    the uniform distribution of the input with a dispersion factor $p
    = 1\%$ and $n=20^3$. Together with the data we display the
    position of the mean $(\mucep\,,\, \Tcep) = ( 0.327 \,,\, 0.026 )$
    (GeV) and the standard deviation $\sigma(T)=0.019$ GeV and
    $\sigma(\mu)=0.0076$ GeV. The $1-$ and $2-\sigma$ approximate
    confidence ellipses are shown also with the principal direction
    ($V_0=0.0013$ GeV and $V_1=0.0138$ GeV are eigenvectors of the
    covariance matrix of the data whose length is given by the square
    root of the corresponding eigenvalue). The confidence ellipses
    drawn at the $1-$ and $2-\sigma$ level are a convenient way to
    visualize the covariance matrix but there are not the true $1-$
    and $2-\sigma$ confidence level.  As the dispersion factor raises,
    the CEPs ``disappear from bellow''. }
\end{figure}

In \reffig{fig:CEP-dist} we present a
scatter plot of the CEPs obtained in this calculation with the
confidence ellipses at $1-\sigma$ and $2-\sigma$. The confidence
ellipses are an approximation of the true $1-\sigma$ and $2-\sigma$
confidence level since the density $\rho(T_{CEP}, \mu_{CEP})$ is not a
Gaussian distribution. These ellipses are just a convenient way to
represent the covariance matrix since the semi-major and semi-minor
axis are the eigenvectors of this matrix.

\begin{figure}[!h]
\includegraphics[width=0.4\textwidth]{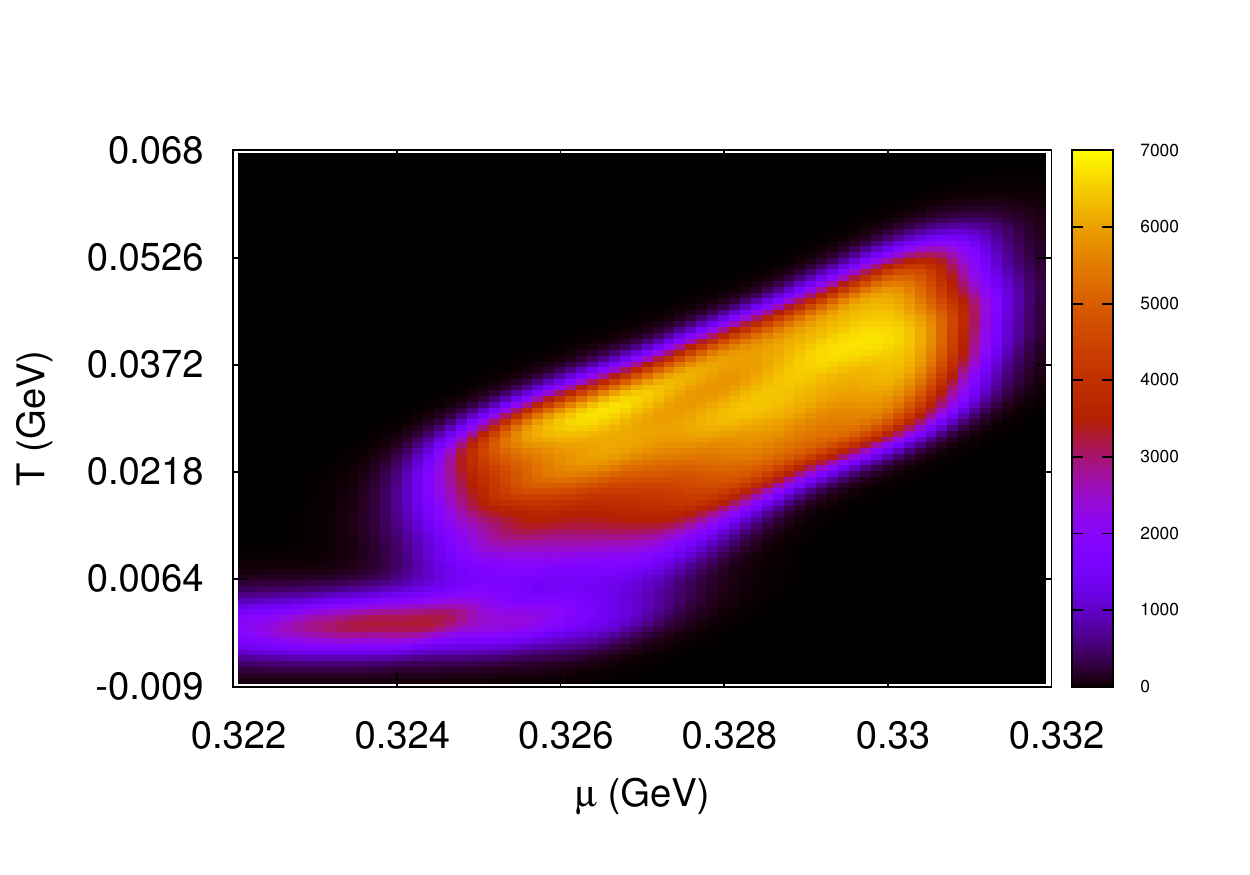}
\caption{ \label{fig:cep-density} Kernel Density Estimate (KDE) of the
  CEP distribution probability in the same simulation condition as
  previously with $p = 1\%$ for the uniform density. The color coded
  z-axis is in GeV$^{-2}$. All parameter sets give rise to a data
  point (notice the points on the $T=0$ line: if the CEP cannot be
  found it is replace by a $T=0$ point (see text)).  }
\end{figure}

Another noticeable result is that with a dispersion as low as $1\%$
(resulting in at most $3$ MeV of variation of the phenomenological
inputs of the model) the CEP scatter plot extends in a large range of
temperature $T_{\textrm{CEP}} \, \in \, [0\,,55]$ MeV and a more
reduced range of chemical potential $\mu_{\textrm{CEP}} \, \in \,
[324\,,332]$ MeV (the standard deviation is of course much
smaller). Furthermore, it exists some parameter sets for which a CEP
does not exist. For all this sets, the phase transition at $T=0$ is a
crossover. To visualize this feature, a point for the nonexistent CEP
is added: its temperature is taken as $T=0$ (the CEPs disappear ``from
below'') and its chemical potential such that $\dd \mu / \dd m = 0$ at
$T=0$\footnote{It is the characteristic crossover chemical potential
  at $T=0$.}. To visualize also the density of the point we represent
in \reffig{fig:cep-density} the probability distribution of the CEP
using the Kernel Density Estimate or KDE (shortly described in
\refapp{app:Kernel density approximation}). The integral on
$\mathbb{R}^2$ of this distribution is one and its dimension is
GeV$^{-2}$.

In \reftab{tab:in-medium-qual-results} the mean and standard deviation
are listed. The ratio of the deviation over the mean of the
temperature may seem low compared to the value of the sensitivity. It
is an artifact coming from the fact that the ``missing CEP'' cannot be
taken into account hence lowering artificially this ratio.
This ratio for the chemical potential is remarkably low.

These results are in agreement with the values of the sensitivities at
vanishing dispersion that we found for the CEP. As already noticed, the
problem is sufficiently linear around the mean value for the
sensitivities calculation to make sense when extrapolated at finite
dispersion.

\begin{table}[h]
  \centering
  \begin{tabular}{|l|r|}
    \hline
    $ \overline{\Tcep}$   & $(0. 0303 \pm 0. 0001)$ (GeV) \\ \hline
    $ \sigma (\Tcep) $     & $(0. 0107 \pm 0.0001)$ (GeV) \\ \hline
    $\sigma(\Tcep)/\overline{\Tcep}$ & $35.25$ $(\%)$ \\ \hline\hline
    $ \overline{\mucep}$  & $(0. 3280 \pm 0.0001)$ (GeV)\\ \hline
    $ \sigma(\mucep)$     & $(0. 0018 \pm 0.0001)$   (GeV)\\ \hline
    $ \sigma(\mucep)/ \overline{\mucep}$ & $0.54 $ $(\%)$\\ \hline
  \end{tabular}
  \caption{ \label{tab:in-medium-qual-results} Results obtained for
    the in-medium prediction, \emph{i.e} concerning the CEP position,
    for the uniform distribution, with $n^3 =
    20^3$, and $p = 1\%$.  This analysis does only take into account
    the CEPs that were find. Thus the presented results are biased. If
    the missing CEPs were used in the analysis, the (relative)
    standard deviation would be even larger.  }
\end{table}

\subsubsection{Correlation of the chiral critical end points with the
  inputs}
\label{ssec:Correlation of the chiral critical end points with the inputs}

Finally, to better find how to constrain the model, it may be of
interest to look at the correlations between the CEP coordinates and
the phenomenological inputs (this has been shortly discussed in previous
sections, for example when we studied variation of the CEP with
respect to the condensate).  The correlations between two quantities
$A$ and $B$ that can be either inputs, parameters or predictions may
be accessed through the correlation coefficient $C_{AB}$ defined as
(see \ccite{Dobaczewski1}):
\begin{equation}
  \label{eq:definition-correlation-coef}
  C_{AB} =  \frac{ \left| \mrm{Covar}(A,B) \right| }  {\sigma(A)
    \sigma(B) } \;,
\end{equation}
where $\mrm{Covar}(A,B)$ is the extra-diagonal coefficient of the
covariance matrix of $A$ and $B$. Explicitly, if $A_i$ and $B_i$ are
the datasets generated by the Monte-Carlo, the
covariance matrix elements are:
\beq
\mrm{Covar}(A,B) = \frac 1 {n - 1}
  \sum_{i = 1}^{n} (A_i - \bar A) (B_i - \bar B)
\eeq
($\bar A$ and $\bar B$ are the means of the corresponding dataset),
$\mrm{Covar}(B,A) = \mrm{Covar}(A,B)$ and the diagonal elements are
simply the variance \textit{e.g.} $\mrm{Covar}(A,A) = \sigma^2(A)$.

\begin{figure*}[!ht]
\begin{center}
\includegraphics[width=0.4\textwidth]{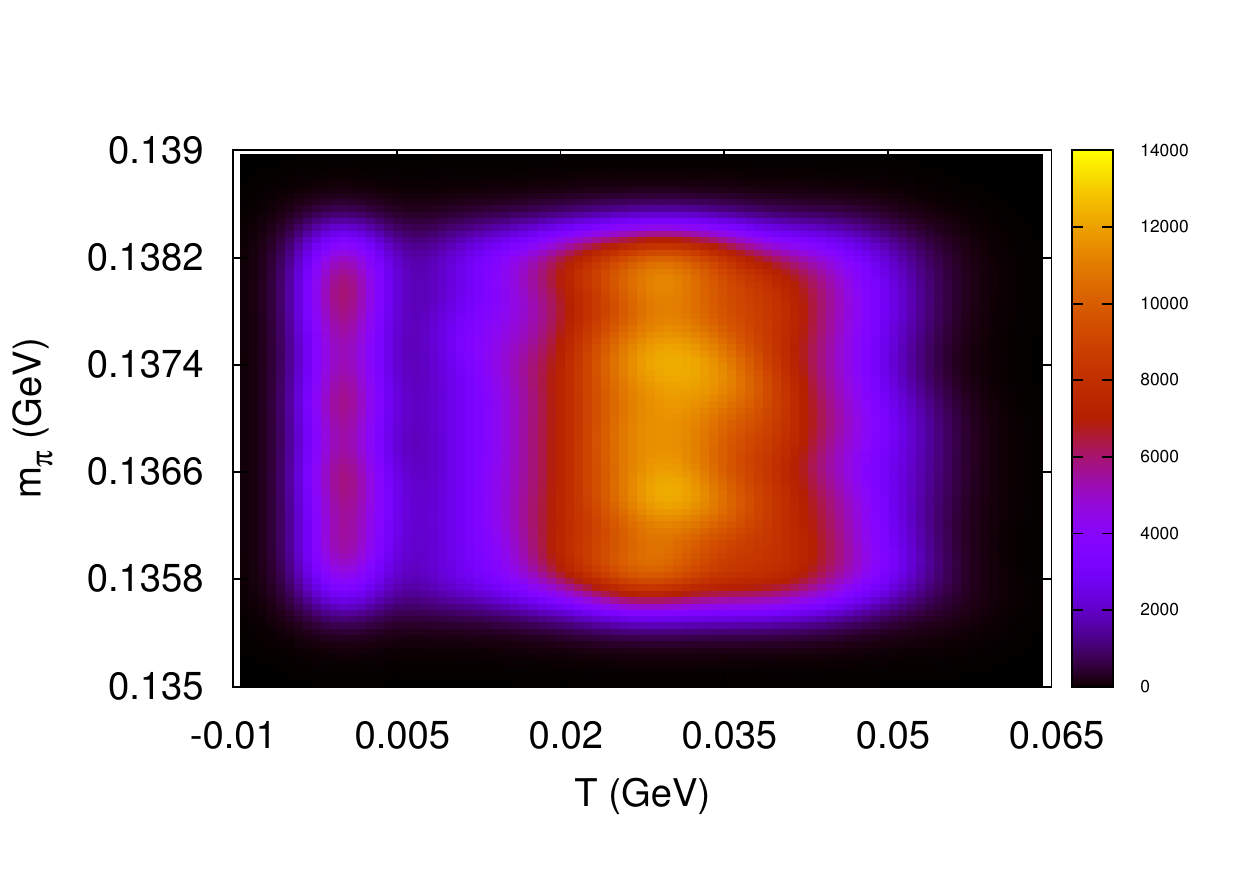}
\includegraphics[width=0.4\textwidth]{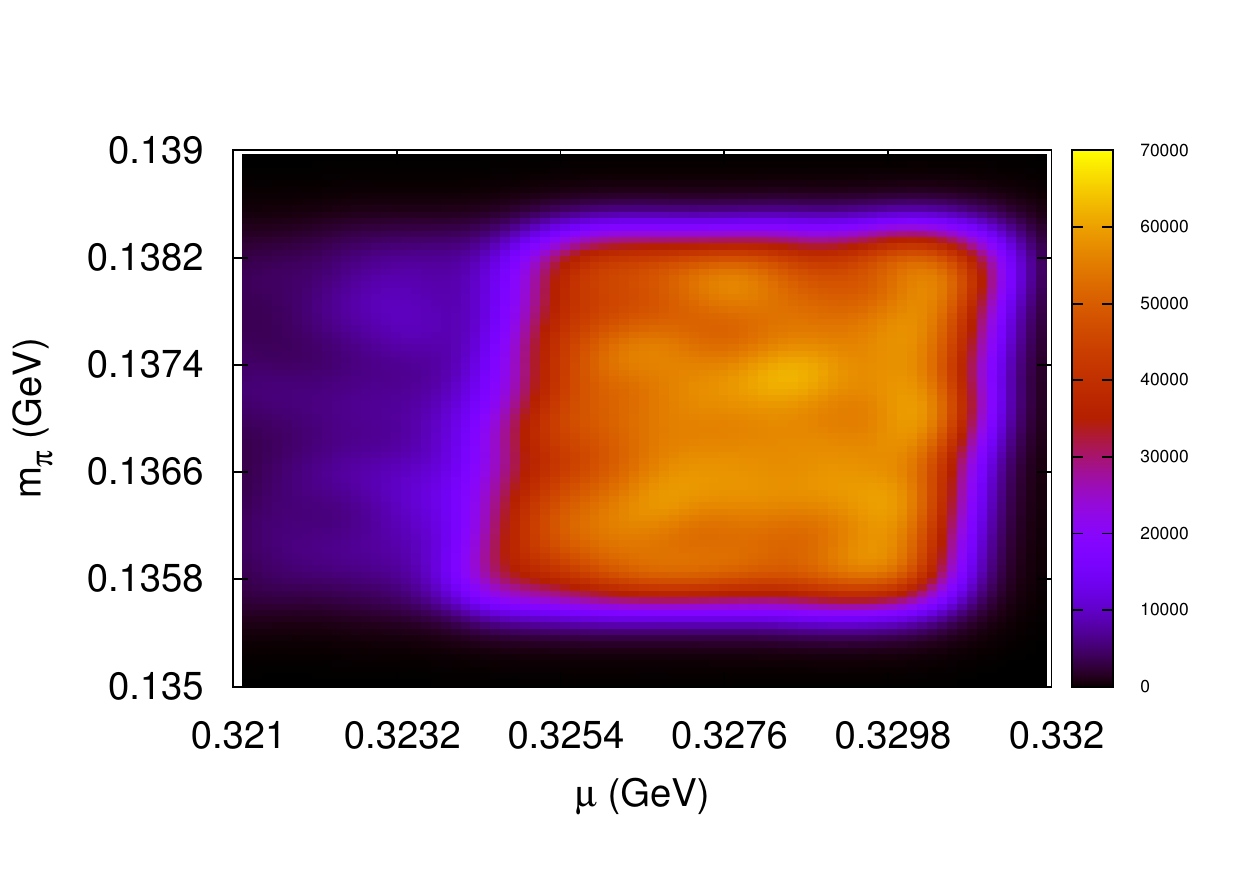}

\includegraphics[width=0.4\textwidth]{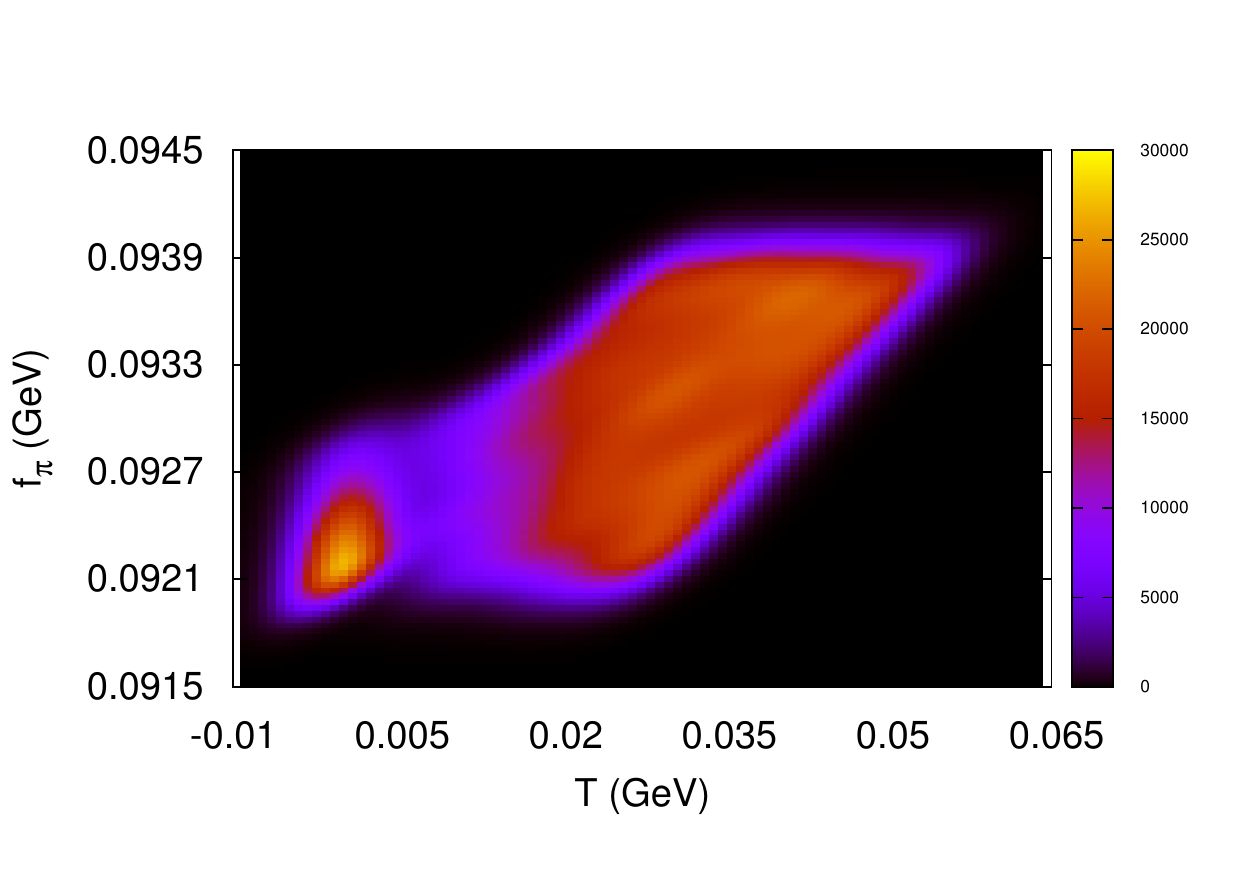}
\includegraphics[width=0.4\textwidth]{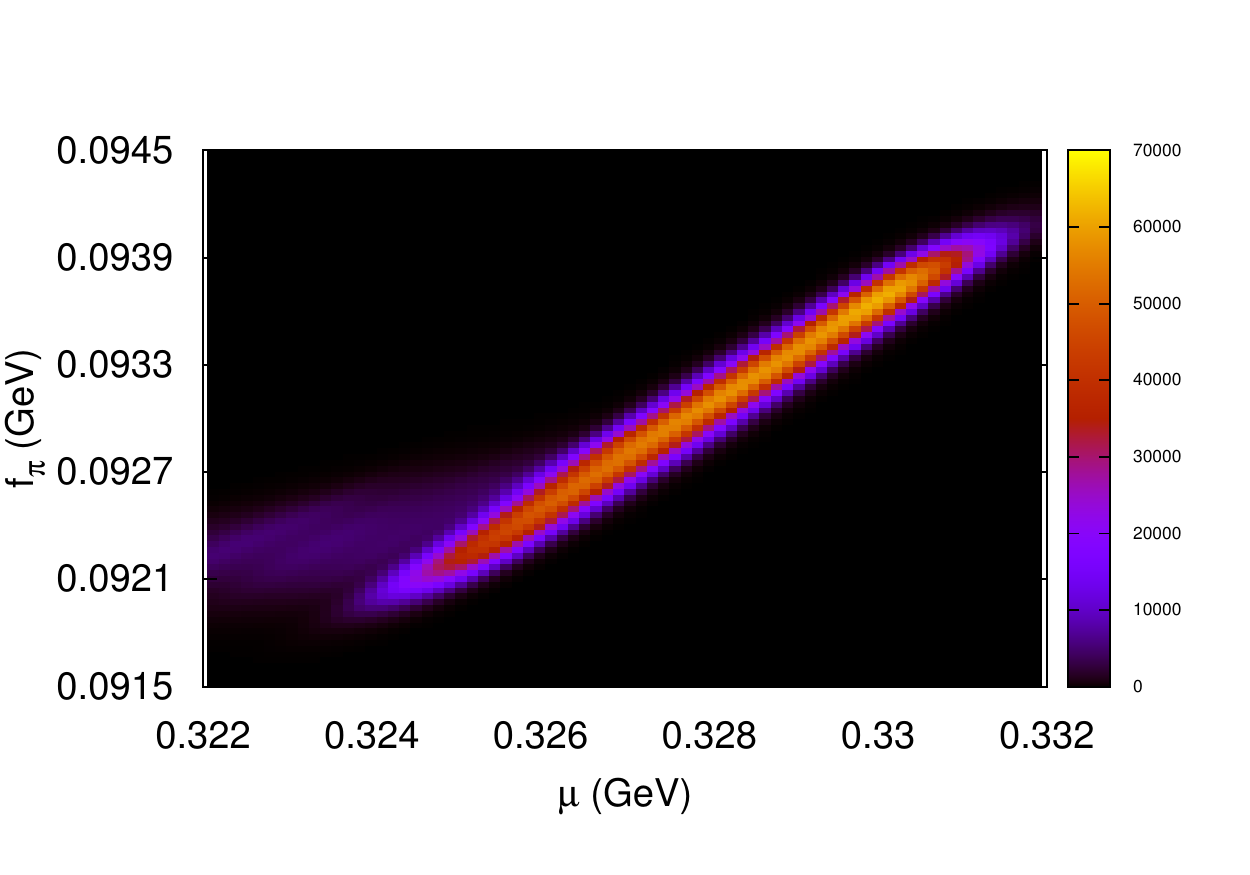}

\includegraphics[width=0.4\textwidth]{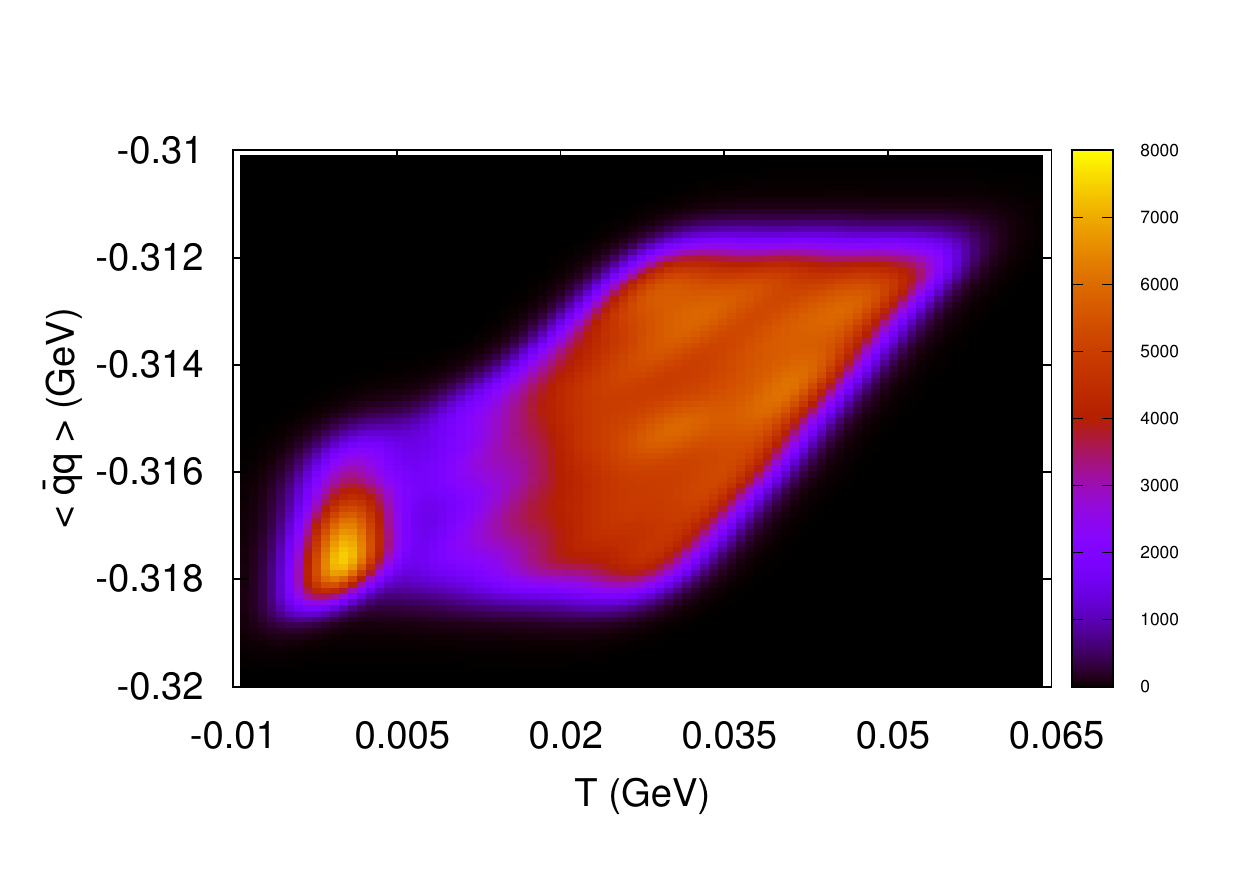}
\includegraphics[width=0.4\textwidth]{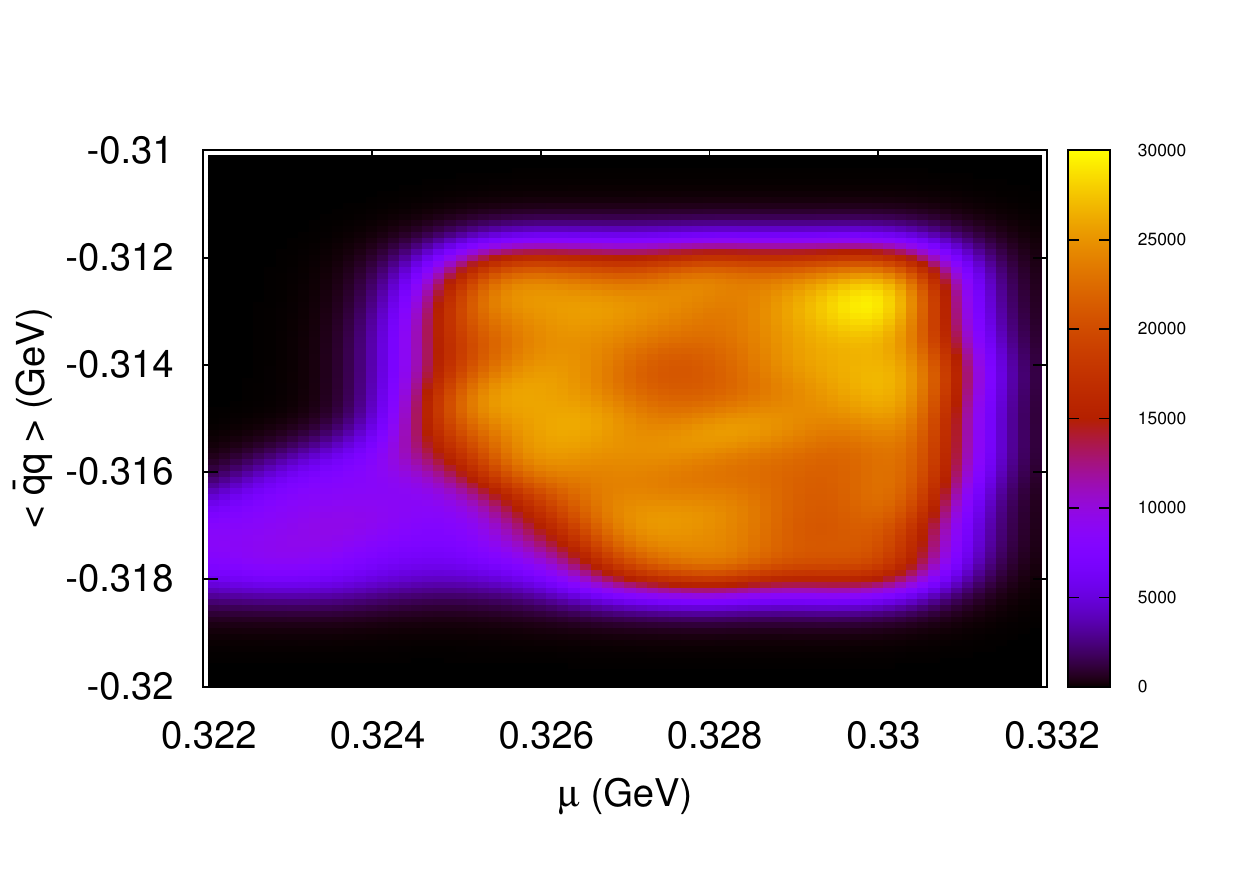}
\end{center}
\caption{ \label{fig:correlations-inputs} Correlation of the
  temperature (left) or the chemical potential (right) with the
  inputs.  Uniform distribution, $p = 1\%$ MeV and $N = 20^3$. To
  represent the correlations we have done a scatter plot of the two
  datasets then reconstructed the density of points with the KDE
  algorithm \refapp{app:Kernel density approximation}. The color
  coded z-axis is then in GeV$^{-2}$.}
\end{figure*}

When this coefficient is close to one it means a strong correlation
between $A$ and $B$ (for example $C_{AA}$ is obviously equal to
$1$). On the contrary a value close to zero means that the random
variable $A$ and $B$ are uncorrelated. To be precise, let us mention
that the coefficient is a good measure of independence of a variable
in the linear case; in the non-linear framework that we are working
in, close to zero correlation may not imply that the two variables are
almost independent. For this reason it is also important to check the
conclusion one can infer from $C_{AB}$ by inspecting the scatter plots
of the datasets $A$ and $B$ (\reffig{fig:correlations-inputs}). Since
we start with a uniform distribution, perfect correlations will result
in a line in the plot where the density of point is constant (and
implies $C_{AB} = 1$) and no correlations will result in a rectangular
shape where the density of points is also constant (and implies
$C_{AB} = 0$).

We will concentrate on the correlations between the CEP coordinates
and the phenomenological inputs or parameters. Indeed since the
predictive power of the model is quite poor it is important to know
which inputs or parameters should be better constrained to restore the
robustness of the CEP prediction. On \reffig{fig:correlations-inputs}
we display correlation plots between the inputs and the CEP
(temperature and chemical potential) with a dispersion of the inputs
of $1\%$ as in the previous plots. To complete this analysis,
\reftab{tab:correl} puts together the values of the correlation
coefficient \refeq{eq:definition-correlation-coef} between the inputs
or the parameters and the CEP coordinates.

We notice that the CEP position is almost uncorrelated to the value of
$m_\pi$ (another manifestation that the Goldstone theorem is a
constraint strong enough to ensure that quantities not involving $m_0$
are largely insensitive to the precise mass of the pion) as can be
seen on the plots (almost a perfect rectangular shape) or in the table
($C$ is below $0.1$ for both $T$ and $\mu$).  Otherwise, there is a
strong correlation of the CEP with $f_\pi$. \\
Let us stress that the plots are richer than the value of the
coefficients. For example, we see that the ``no CEP'' points
(represented as zero temperature points) correlates with
low values of $f_\pi$ or the condensate.
The only difference between $T$ and $\mu$ correlation coefficients with
inputs concerns the condensate. The temperature is 4 times more
correlated with it than the chemical potential. As already mentioned,
for the sigma mass and when we vary the condensate, this kind of relation
could not have been easily found based on physical arguments,
illustrating the power of the correlation analysis and the unexpected
behavior of the inverse problem. This difference between $T$ and $\mu$
correlations may be one of the reason why the temperature prediction is
less robust (more dispersed) than the chemical potential one.

\begin{table}
  \begin{center}
    \begin{tabular}{|l|c|c|c|}
        \hline
        Correlations of & & $T_{CEP}$ & $\mu_{CEP}$  \\
        \hline
        \multirow{3}{*}{with inputs} 
        & $m_\pi$   &  $0.021$ &  $0.123$ 
        \\ \cline{2-4} 
        & $f_\pi$   &  $0.646$ &  $0.987$ 
        \\ \cline{2-4}
        & $\qbarq$ &  $0.591$ &  $0.130$ 
        \\ \hline        
        \multirow{3}{*}{with parameters} 
        & $m_0$   &  $0.797$ &  $0.494$ 
        \\ \cline{2-4} 
        & $\Lambda$    &  $0.933$ &  $0.445$
        \\ \cline{2-4}
        & $G\Lambda^2$ &  $0.975$ &  $0.686$ 
        \\ \hline
      \end{tabular}
   \end{center}

  \caption{
    \label{tab:correl}
    Correlation coefficients between the inputs or the model
    parameters with the temperature and the chemical potential at CEP
    with uniform distribution, $p = 1 \%$, and $n = 20^3$.
  }
\end{table}

%
%

\section{Conclusion}

In this paper, the effects of infinitesimal variations of the inputs
used to fix the parameters of the NJL model were systematically
studied using the sensitivity parameter.

This work is a benchmark for more realistic studies but we already
shown how poorly the CEP temperature is predicted by the model (if it
can at all be predicted when considering finite variations).

The great advantage of the sensitivity is to quantitatively and
systematically assess the information on how a quantity is reasonably
predicted in a model without being obligated to vary by hand the value
of the parameter to try to assess the stability of a
prediction. Besides, the inverse problem analysis is done at constant
vacuum phenomenology.

Along the way we have illustrated how powerful the tools from the inverse
problem theory can be, revealing non trivial relations between inputs
and predictions, for example when studying the sigma mass (whose physics is
well understood within this model) or when softening the constraint on
the condensate value (where we found a less trivial behavior). 

The simulations requiring a large number of samples, we also show as
an exercise in \refapp{app:inverse-pb} an exact solution of the
inverse problem and, \refapp{app:cep-calc}, a fast algorithm to
compute the CEP by pushing as much as possible the analytic
calculation (in particular using no numerical derivatives that are
known to be quickly badly behaved). In more realistic models, the
former certainly won't be possible anymore, but our algorithm can
be easily generalized. 

The general conclusion is that with parameters fitted to the in-vacuum
quantities, the intuition dictates that in-medium prediction may be
less safe (even if it is the usual framework in the literature). What
we have shown here is a way to give a quantitative statement about
this intuition. We also show that for the case of the chemical
potential it seems to go opposite to the intuition.  With this work we
suggest that for more realistic models such a sensitivity criterion
(together with correlation analysis) may help to know in which
direction the model should be refined and constrained to have better
predictions. 

One noticeable result that we will investigate further is that the
chemical potential of the CEP is remarkably stable. Of course the
vector interaction will change this value (the vector interaction essentially
shifts the chemical potential) but we conjecture and we will check it
in a next publication if the temperature still remains unpredictable.

From this remarkable stability we learn even with this simplified
model that if the CEP is shown to exist (\textit{e.g.} observationally if one
has a proof that the transition is first order at zero temperature),
the contribution of the scalar interaction will contribute to strongly
constraint its chemical potential coordinate but the prediction of its
temperature is not possible in this framework. It shows that one has
to add other relevant physical mechanism for this prediction. We
suppose that, in the PNJL model which add finite temperature
constraints, this problem would be lessened (but not completely
solved) since gluonic effect have an important role to fix the
temperature.

We will also test if adding in medium inputs will indeed stabilize
enough the prediction for it to be meaningful.  The problem being of
course that there is few finite density experimental/theoretical
constraints to QCD. We will also study if even a very weak constraint
(for example one coming from compact star phenomenology where
experimental uncertainties are large) is able to stabilize the CEP and
hence showing that effective models can be trust as a useful tool to
study QCD where for example LQCD cannot reach.

Let us conclude on a more general note. The CEP position in this
simple model is rather low in the phase diagram (small $T$, high
$\mu$), at the bottom of the chiral crossover and 1st order transition
line. It may be related to the fact that its sensitivity is rather
low: when varying parameters, the CEP essentially follows the
transition line that is rather steep in this part of the phase diagram
hence its value do not change a lot.

On the contrary LQCD or first experimental evidences
\ccite{Lacey:2014wqa} favor a much lower value of the chemical
potential and a higher value of the temperature.

We already have shown \ccite{Costa2010} (Fig. 8) that with the SU(3)
NJL model with a Polyakov loop (that take into account a static gauge
field) the CEP is considerably higher in the phase diagram.  We also
shown that in order to put it even closer to the zero chemical
potential axis (in better agreement with the aforementioned evidences)
we must either force the strange mass to a very low, nonphysical value
(Fig. 8 of {\ccite{Costa2010}) as it is well known (when all light
  quark masses are low the transition at zero chemical potential is
  first order) or impose a large value of the t'Hooft coupling
  constant (Fig. 9 of \ccite{Costa2010}) hence destroying the correct
  magnitude of the eta - eta' meson mass difference. If this evidences
  are confirmed it will be a priori difficult to get the correct
  position of the CEP with a correct description of the vacuum. It is
  the great advantage of this framework to be able to systematically
  study parametrization of the model that obey a given set of
  constraints (contrarily to the above mention work where we vary
  independently parameters without inspection of the induced
  phenomenology). In future works we will use these tools to check if
  a parametrization exists that can reproduce both reasonable vacuum
  mesonic spectrum and in-medium CEP position.  If not, it may be an
  indication that some important effects are missing in the model
  either higher order correction, back reaction mechanism or even new
  mechanism absent of the model like dynamical contribution of the
  gluonic sector.

%
%

\begin{appendix}

\section{NJL model and its analytic inversion}
\label{app:inverse-pb}

The NJL model we consider, whose Lagrangian is given by
\refeq{eq:NJL-Lagrangian}, has three parameters $m_0$, $\Lambda$ and
$G$ that are fitted to the values of the pion mass, $m_\pi$, the pion
decay constant $f_\pi$, and the quark condensate $\qbarq$.

Let us recall the system of equations for the inverse problem we
obtained in the text (\refeq{eq:mpi}, \refeq{eq:fpi} and
\refeq{eq:cond}). When $(m_\pi,\ f_\pi, \qbarq)$ are fixed to their
phenomenological values, one has to solved for the parameters $\Lambda$,
$m_0$ and $G$ the system:
\beq
\label{eq:app-inverse-problem-sys-njl-1}
    m_\pi^2   & = &  - \frac{m_0}{m} \frac{1}{4 i G N_c N_f I_2(0)} \;, \\
\label{eq:app-inverse-problem-sys-njl-2}
    f_\pi^2   & = &  -4i N_c m^2 I_2(0)  \;, \\
\label{eq:app-inverse-problem-sys-njl-3}
    \qbarq   & = & \frac{m_0 - m}{2G}\;,
\eeq
together with the equation for the mass $m$ (\refeq{eq:gap-equation}): 
\begin{equation}
  \label{eq:app-gap-equation}
  m_0 - m + 8iGN_cN_f m I_1 = 0 \;.
\end{equation}

\subsection{Reduction of the system}
The idea to reduce the system is to use adimensional quantities hence
we conveniently rewrite \refeq{eq:I1} and \refeq{eq:I2}:
\beq
I_1 &=& -i \Lambda^2 i_1(m/\Lambda), \nonumber \\
\label{eq:app-definition-i1}
\mbox{where } i_1 (x) &\equiv& 
                       \int^1 \frac{\dd^3 p }{(2\pi)^3} 
                       \frac{1}{2\sqrt{p^2+x^2}}  \;,
\eeq
and
\beq
I_2(0) &=& \frac i 4 i_2(m/\Lambda), \nonumber \\
\mbox{where } i_2(x) &\equiv& 
                    \int^1 \frac{\dd^3 p }{(2\pi)^3} 
                    \frac{-1}{ \left({p^2+x^2}\right)^{3/2}} \;.
\label{eq:app-definition-i2}
\eeq

The  integrals $i_1$ and $i_2$ can be computed analytically:
\beq
\label{eq:app-integral-i1-analytic}
i_1(x) & = & \frac{1}{8\pi^2} \left[ 
\Lambda_E + x^2 \log \left( \frac{x}{1+ \Lambda_E} \right) \right]\;, \\
i_2(x) &=&  \frac{1}{2\pi^2} \left[ 
\frac{1}{\Lambda_E} +  \log \left( \frac{x}{1 + \Lambda_E} \right) \right]\;,
\eeq
with $\Lambda_E = \sqrt{x^2 + 1}$.

To solve the system, the scale $\Lambda$ is used to make the variables
dimensionless: 
\[
x = m / \Lambda  
\quad \textrm{and}\quad 
x_0 =  m_0 / \Lambda 
\]
With these variables, the system reads:
\beq
\label{eq:app-inverse-problem-sys-njl-1-new}
\frac{m_\pi^2}{\Lambda^2} &=& \frac{x_0}{x} \frac{1}{G\Lambda^2 N_c N_f i_2(x)} \;,\\
\label{eq:app-inverse-problem-sys-njl-2-new}
\frac{f_\pi^2}{\Lambda^2} &=& N_c x^2 i_2(x) \;,\\
\label{eq:app-inverse-problem-sys-njl-3-new}
\frac{\qbarq}{\Lambda^3} &=& \frac{x_0 - x}{2 G \Lambda^2} \;,
\eeq
and the gap equation becomes:
\begin{equation}
0 = x_0 - x + 8 G \Lambda^2 N_c N_f x i_1(x) \;.
\end{equation}
This equation is automatically solved if:
\begin{equation}
  \label{eq:app-equation-for-Gamma}
G\Lambda^2 = \frac{x-x_0}{8 N_c N_f x i_1(x)} \;.
\end{equation}
This form for $G\Lambda^2$ can be plugged in
\refeq{eq:app-inverse-problem-sys-njl-2-new} and
\refeq{eq:app-inverse-problem-sys-njl-3-new}. Introducing another new
variable:
\[
\delta = \frac{x - x_0}{x_0} \;,
\]
the system now reads:
\beq
\label{eq:app-Sys-1}
\frac{m_\pi^2}{\Lambda^2} &=& \frac{8 i_1(x)}{\delta i_2(x)} \;,\\
\label{eq:app-Sys-2}
\frac{f_\pi^2}{\Lambda^2} &=& N_c x^2 i_2(x) \;,\\
\label{eq:app-Sys-3}
\frac{\qbarq}{\Lambda^3} &=& -4 N_c N_f x i_1(x) \;.
\eeq

With this last form, we can solve it by first calculating the ratio:
\begin{equation}
\label{eq:app-def-alpha}
\alpha = \frac{f_\pi^3}{\qbarq} ,
\end{equation}
which is a phenomenological constant independent of $\Lambda$ and hence, by
taking the quotient between \refeq{eq:app-Sys-2} and
\refeq{eq:app-Sys-3}, $x$ is the solution of one equation of one unknown
$\alpha$:
\begin{equation}
  \label{eq:app-decomp-sys-1}
G_\alpha(x) = 0 \;,
\end{equation}
with: 
\begin{equation}
  G_\alpha(x) = \frac{\sqrt{N_c}}{4N_f} \frac{x^2 (i_2(x))^{3/2}}{i_1(x)} + \alpha  \;.
\end{equation}

Once \refeq{eq:app-decomp-sys-1} is solved, we can compute $\delta$ by
computing the ratio $f_\pi^2/m_\pi^2$ leading to:
\begin{equation}
  \label{eq:app-decomp-sys-2}
  \delta = \frac{f_\pi^2}{m_\pi^2} \frac{8i_1(x)}{N_c(xi_2(x))^2} \;.
\end{equation}

Finally, the system has just to be re-scaled:
\begin{equation}
  \label{eq:app-decomp-sys-3}
  \Lambda = \frac{f_\pi}{x \sqrt{N_ci_2(x)}} \;.
\end{equation}
The values of $m_0$ and $G$ can now be calculated:
\beq
\label{eq:app-m0-equation}
m_0 &=& \Lambda \frac{x}{\delta + 1} \;,\\
\label{eq:app-G-equation}
G &=& \frac{1}{\Lambda^2} \frac{x - x_0}{ 8  N_c  N_f x  i_1(x)} \;.
\eeq

\subsection{Solution for $x$}

\begin{figure}[!h]
\includegraphics[width=0.48\textwidth]{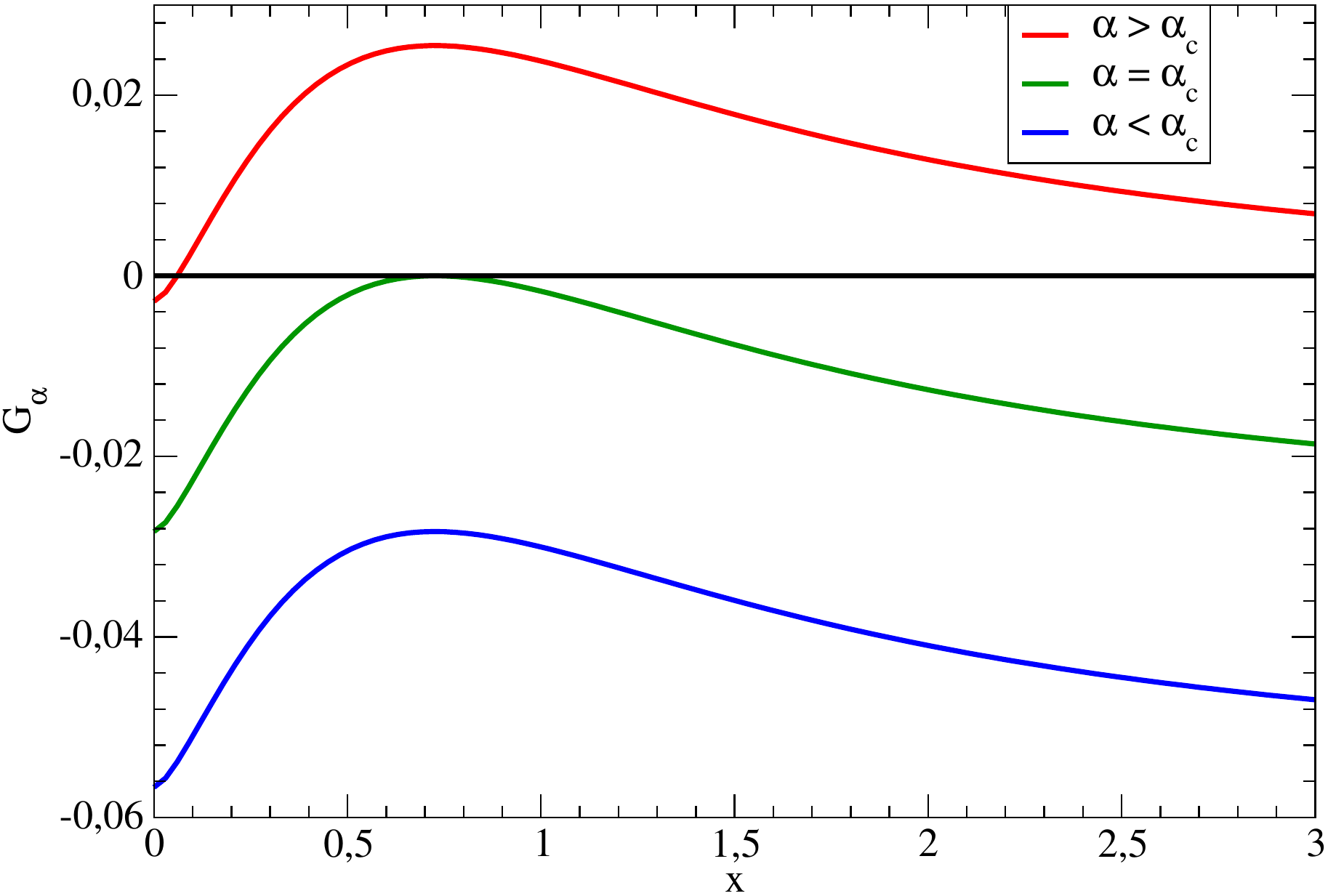}
\caption{Function $G_\alpha(x)$ for three typical values of the
  parameter $\alpha$ (always negative for physical values of the
  phenomenological inputs): \newline (top) $0 > \alpha > \alpha_c$: two
  solutions; (middle) $\alpha = \alpha_c$: one degenerate solution at
  maximum; (bottom) $\alpha < \alpha_c$: no solution.  If $\alpha >
  \alpha_c$ the system has two solutions: one is physical, for $x <
  x_{\mrm{max}}$, while the other is not, for $x>x_{\mrm{max}}$, where
  $x_{\mrm{max}}$ is the abscissa of the maximum of $G_\alpha$
  function. }
\label{fig:app-Galpha} 
\end{figure}

Provided that none of the parameters is zero, the system given by
\refeqs{eq:app-Sys-1}{eq:app-Sys-2}{eq:app-Sys-3} is equivalent to
\refeq{eq:app-decomp-sys-1}). This equation has to be solved for
$x$, and depending on the value of $\alpha$ given
\refeq{eq:app-def-alpha}, the system may be solved or not.  In
\reffig{fig:app-Galpha}, the function $G_\alpha$ is plotted for three
values of $\alpha$. Calling $x_{\mrm{max}}$ the abscissa such as
$G_\alpha$ is maximum, then we see that \refeq{eq:app-decomp-sys-1}
has solutions for each $\alpha > \alpha_c$, where $\alpha_c$ is
defined as the $\alpha$ for which the value of the function $G_\alpha$
at $x_{\mrm{max}}$ is zero:
\begin{equation}
  \label{eq:app-definition-of-alpha-c}
  \alpha_c \quad \Leftrightarrow \quad G_{\alpha_c}(x_\mrm{max}) = 0 \;.
\end{equation}
Since the abscissa of the maximum value of $G_{\alpha}(x)$ is
independent of $\alpha$, the critical value is a numerical constant
$\alpha_c = -0.0283275$ and has just to be calculated once. For all
values of $\alpha$ that respect $\alpha > \alpha_c$ the system has two
solutions. The first one is numerically found by any bracketing
algorithm which is looking for a root in $[0 , x_{\mrm{max}}]$. This
root corresponds to the physical dressed mass $x = m/\Lambda <
x_{\mrm{max}}$.
The second root corresponds to an nonphysical mass $m/\Lambda >
x_{\mrm{max}}$ (it is another way to see the phenomenon described in
\ccite{NJLrev_buballa}, Fig. 2.6).
Nevertheless, as a complement, we
present a way to find it.  It is possible to calculate an asymptote of
$G_{\alpha}(x)$ at $x\to +\infty$ defined by:
\begin{equation}
  G^{\infty}_{\alpha}(x) = \sqrt{\frac{N_c}{3}} \frac{x^{-3/2}}{4\pi N_f} + \alpha \;.
\end{equation}
The solution of $G^{\infty}_{\alpha}(x) = 0$ is analytic and reads:
\begin{equation}
  x^{\infty} = - \left(\frac{N_c}{24}\right)^{1/3} (\alpha \pi N_f)^{-2/3} \;.
\end{equation}
The nonphysical root can be searched in the interval \newline \mbox{$[x_{\mrm{max}} ,
x^{\infty}]$}.

\section{Fast algorithm for the chiral critical end
  point calculation}
\label{app:cep-calc}

To study the in-medium properties of the NJL model, one has to
generalize the in-vacuum gap equation \refeq{eq:app-gap-equation} for
finite temperature and finite chemical potential. The gap equation to
be solved is:
 \begin{equation}
  \label{eq:finite-T-mu-MFE}
  g_m(m,T,\mu ) = 0 \;,
\end{equation}
where:
\begin{equation}
  \label{eq:finite-T-mu-gm}
  g_m = m_0 - m + 8 G N_c N_f m  
         \left[ i I_1(m,\Lambda) - I_\beta(m,T,\mu) \right]  \;,
\end{equation}
with $I_1$ being the integral (\ref{eq:app-definition-i1})
and $I_\beta$ the integral defined as:
\begin{equation}
  \label{eq:integral-I-beta}
  I_\beta = \int^{\infty} \frac{\dd^3 p }{(2\pi)^3} \frac{1}{2E_p} 
           \left[ f(p) + \bar{f}(p) \right]  \;,
\end{equation}  
where $E_p =\sqrt{p^2 + m^2}$ and $f$ and $\bar f$ are the Fermi-Dirac 
distribution for quarks and antiquarks respectively
\begin{align}
  f(p) &= \frac{1}{1+\exp\left(\beta( E_p - \mu )  \right)} \nonumber \;,\\
  \bar f(p) &= \frac{1}{1+\exp\left(\beta(E_p + \mu)  \right)} \nonumber \;,
\end{align}
with $\beta = T^{-1}$ and $k_B = 1$.
The solution of \refeq{eq:finite-T-mu-MFE} is plotted 
as a function of the chemical potential for different temperatures 
on \reffig{fig:m-of-Tmu}. 
\begin{figure}[t]
\includegraphics[width=0.48\textwidth]{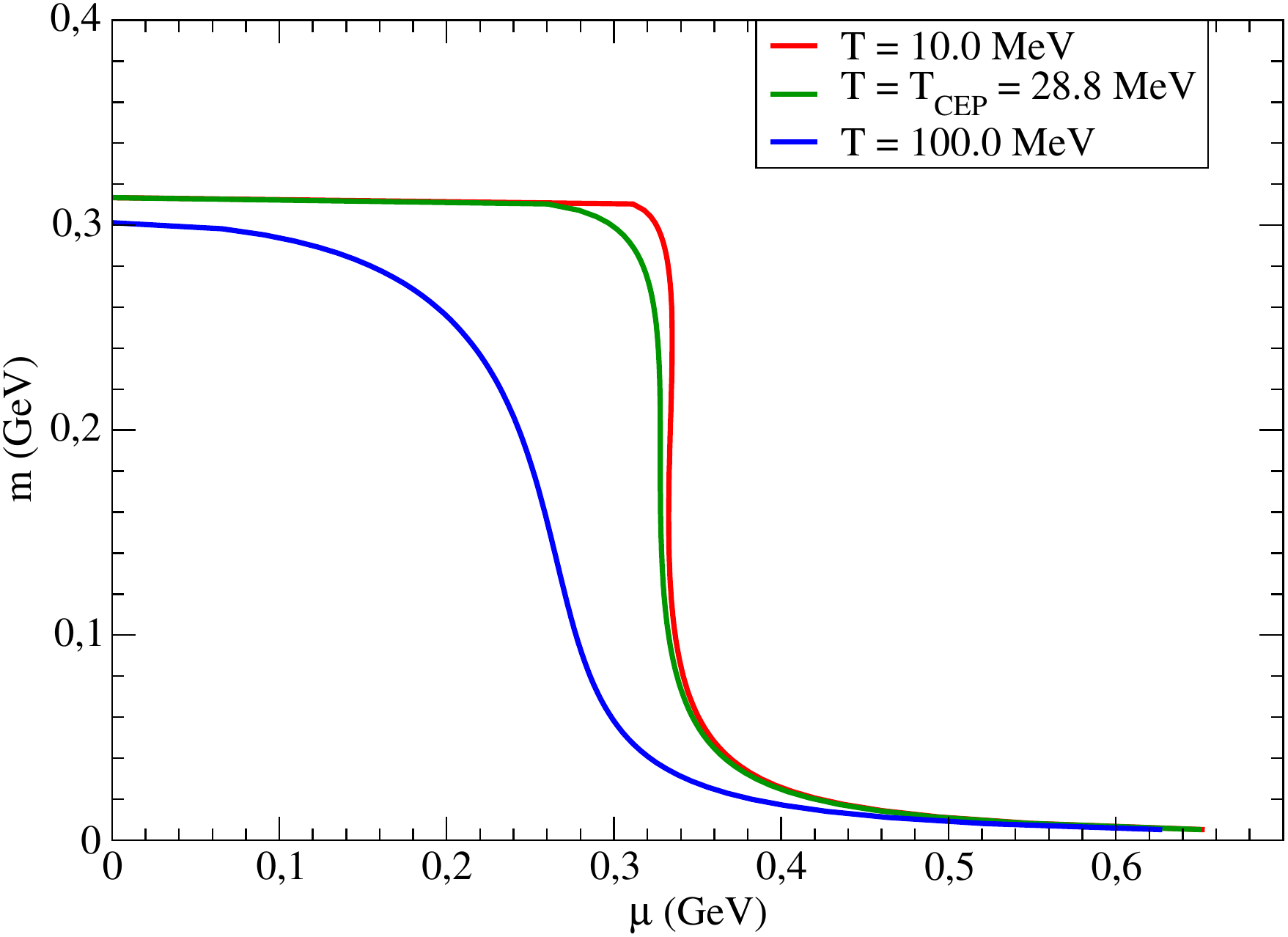}
\caption{\label{fig:m-of-Tmu} In medium dressed quark mass for the
  three typical cases described in text.  }
\end{figure}
It displays three different behaviors:\\
(i) For $T > T_{\mrm{CEP}}$, there is a single solution $m(T,\mu)$
characteristic of a cross-over transition between a chirally
broken hadronic phase ($\qbarq \neq 0$) and an almost chirally
symmetric phase ($\qbarq \simeq 0$);\\
(ii) For $T < T_{\mrm{CEP}}$, it exists, in a range of chemical potential,
three solutions for $m$, characteristic of a first order chiral phase
transition with stable, metastable, and unstable solutions;\\
(iii) For $T = T_{\mrm{CEP}}$, it exists a unique solution $m$ but if $\mu =
\mu_{\mrm{CEP}}$, then the tangent of $m(T,\mu)$ is infinite: at the
CEP, the phase transition is of second order.

Then, to compute the CEP coordinates, one has to compute the
temperature and the chemical potential where $m$ has a unique infinite
tangent (strictly speaking one should work with the order parameter,
namely the quark condensate, but it is equivalent and easier to work
with the quark mass):
\begin{equation}
  \left. \frac{\dd m}{\dd \mu} \right|_{T=\Tcep} = 
       + \infty \quad \Leftrightarrow \quad \left.
               \frac{\dd \mu }{\dd m}\right|_{T=\Tcep} = 0 \;,
\end{equation}
where the function $\mu(m,T)$ is an implicit solution of\\
\mbox{$g_m(m,T,\mu(m,T)) = 0$} (this function is easier to work with since it
is always single valued). Because of the unicity of the infinite
tangent at the CEP, the latter is also an inflection point, meaning
that finding the CEP coordinates means to solve the following system 
of equations:
\begin{align}
  \label{eq:sys-CEP-mfe}
  g_m(m_{\mrm{CEP}},T_{\mrm{CEP}},\mu_{\mrm{CEP}}) &= 0 \; , \\
  \label{eq:sys-CEP-derivative}
  \left.\frac{\dd \mu}{\dd m}\right|_T (m_{\mrm{CEP}},T_{\mrm{CEP}},\mu_{\mrm{CEP}}) 
                          &= 0 \; , \\
  \label{eq:sys-CEP-inflexion}
  \left.\frac{\dd^2 \mu}{\dd m^2}\right|_T (m_{\mrm{CEP}},T_{\mrm{CEP}},\mu_{\mrm{CEP}}) 
                          &= 0 \;.
\end{align}

The traditional way to compute the CEP is to solve numerically 
\refeq{eq:sys-CEP-mfe}, then find the maximum of $\dd \mu / \dd m$ for 
any $T$, and then find $T$ such as the value of $\dd \mu / \dd m$ at this 
temperature is zero. This method works quite well if the derivatives are not 
calculated numerically, and if adequate initial values are given to the 
algorithm. Using this method, the time required to compute the CEP position 
is a fraction of second. It the following we show how this calculation can
be accelerated.

\subsection{Rewriting of the gap equation}

We introduce the new variables 
\[
\sigma = \beta \mu \qquad \textrm{and} \qquad x = \beta m \;.
\] 
Let us stress that the previous definition of the dimensionless
parameters like the mass $x$ was done with the scale $\Lambda$ and now
we use $1/\beta$: we keep the same name since there is no possibility
of confusion. Renaming $p\to\beta p$ in the integral
(\ref{eq:integral-I-beta}), we can write:
\begin{equation}
  \label{eq:app-integrales-I-beta-et-i-beta}
  I_\beta ( m , T , \mu ) = T^2 i_\beta ( \sigma , x ) \;,
\end{equation}
where $i_\beta$ is given by
\begin{equation}
  \label{eq:app-definition-i-beta}
 i_\beta(\sigma,x) = 
              \int^\infty \frac{\dd^3 p }{(2\pi)^3} \frac{1}{2E} 
                          \left[ f(p) + \bar f (p) \right] \;,
\end{equation}
with $E = \sqrt{p^2 + x^2}$, with $f$ and $\bar f$ the Fermi-Dirac
distributions $f = [ 1 + \exp(E\pm\sigma)]^{-1}$.

The mean field equation (\ref{eq:sys-CEP-mfe}) can be rewritten using 
the new variables $x_0 = \beta m_0$, $\gamma = GT^2$, and 
$\lambda = \beta\Lambda$:
\begin{equation}
  \label{eq:app-mfe-for-CEP-calc}
  0 = x_0 - x - 8 \gamma N_c N_f x 
  \left[ i_\beta(\sigma,x) - \lambda^2i_1 (x/\lambda) \right] \;.
\end{equation}
If one introduces the variable $\eta = x / \lambda$, then the previous 
equation can be written:
\begin{equation}
  8 \gamma N_c N_f \eta^2 i_1 (\eta^{-1}) = 
          8 \gamma N_c N_f \frac{i_\beta(\sigma,x)}{x^2} 
                        - \frac{x_0}{x^3} + x^{-2} \;.
\end{equation}

Using the definitions of the new variables, we can compute:
\begin{equation}
  \frac{1}{\gamma x^2} = 
                  \frac{\eta^2}{G\Lambda^2} 
                          \quad \mrm{and} \quad 
  \frac{x_0}{\gamma x^3} = 
                  \eta^3 \frac{m_0}{\Lambda} \frac{1}{G\Lambda^2} \;,
\end{equation}
and, introducing:
\begin{equation}
  \label{eq:app-definition-of-coef-a-and-b}
  a = ( 8 G\Lambda^2 N_c N_f )^{-1} 
                 \quad \mrm{and} \quad 
  b = \frac {m_0} \Lambda \;,
\end{equation}
the mean field equation becomes:
\begin{equation}
  \label{eq:app-mfe-change-var}
  a \eta^2 ( b \eta -1) + \eta^2 i_1 ( \eta^{-1} ) = 
              \frac{i_\beta(\sigma , x )}{x^2} \;.
\end{equation}

We introduce the function $F(\eta)$:
\begin{equation}
  \label{eq:definition-F-eta}
  F(\eta) = a \eta^2 ( b \eta -1) + \eta^2 i_1 ( \eta^{-1} ) \;,
\end{equation}
and the function $Z(\sigma ,x)$:
\begin{equation}
  \label{eq:definition-Z-sigma-x}
  Z(\sigma ,x) =  \frac{i_\beta(\sigma , x )}{x^2} \;,
\end{equation}
Such that \refeq{eq:app-mfe-change-var} simply reads:
\begin{equation}
  F(\eta) =   Z(\sigma ,x) \;.
\end{equation}

Since integral (\ref{eq:app-definition-i-beta}) is given by a
numerical integral, calculating its inverse is time consuming. On the
contrary $i_1(x)$ is analytical and $F^{-1}(\eta)$ can be efficiently
computed with a simple root polishing algorithm without numerical
integration. At fixed $\sigma$ and $x$, the solution $\eta$ of the
mean field equation is:
\begin{equation}
  \label{eq:app-inverse-of-F-eta-solution}
  \eta_{\mrm{MFE}}(\sigma, x) = F^{-1} \circ Z ( \sigma , x ) \;,
\end{equation}
where $\eta_{\mrm{MFE}}$ is then the $\eta$ that solves the mean field equation, 
from which one can compute the mass, the temperature, and the chemical 
potential using:
\begin{equation}
  m = \frac{\Lambda}{\eta_{\mrm{MFE}}} \;\; ; \quad 
  T = \frac{\Lambda}{x\eta_{\mrm{MFE}}} \;\; ; \quad 
  \mu = \frac{\Lambda \sigma}{x \eta_{\mrm{MFE}}} \;.
\end{equation}

One may be surprised to work with these variables but, in fact, using
these variables is equivalent to choose a trajectory on the surface
defined by:
\begin{equation}
  g_m(m,T,\mu) = 0 \;.
\end{equation}
For example, at fixed $\sigma$, computing $\eta(x)$ and then 
\mbox{$m(x) = \Lambda / \eta(x)$} is equivalent to compute $g_m(m,T,\mu) = 0$ 
with the constraints $\sigma = \mrm{cst} = \mu/T$.

The parametric curve 
$\left\{m=m(x)\, , \, \mu = \Lambda\sigma / [x\eta(x)]\right\}$ 
at $\sigma$ fixed is simply the solution of:
\begin{equation}
\label{eq:app-gap-parametric}
g_m \left( m(\mu) , T = \frac{\mu}{\sigma} , \mu\right) = 0 \;,
\end{equation}
{\it i.e.}, it is the solution of the mean field equation on lines 
$T = \mu / \sigma$ in the $(T,\mu)$ plane.

We have to be careful in the following since the CEP is defined as the
point where $m$ has an infinite derivative with respect to the
thermodynamical parameters $T$ and $\mu$.  In the light of
\refeq{eq:app-gap-parametric}, computing the CEP coordinates is
equivalent to solve the system given by 
\refeqs{eq:sys-CEP-mfe}{eq:sys-CEP-derivative}{eq:sys-CEP-inflexion} at a
fixed $\sigma$. As a remark we notice that working at fixed $x$ gives
simpler equations but one can check that the solution found will not
be the CEP. Indeed the trajectory followed on the surface $S = \{
g_m(m,T,\mu) = 0 \}$ is not trivial and the link between this solution
and the CEP is not obvious.

\subsection{Finding the CEP}

Using the previous notations, the system can be rewritten as:
\begin{align}
  \label{eq:app-sys-CEP-new-MFE}
  F(\eta) &= Z(\sigma , x) \;, \\
  \label{eq:app-sys-CEP-new-derivative}
  \left. \frac{\dd \mu}{\dd m} \right|_{\sigma} &= 0 \;, \\
  \label{eq:app-sys-CEP-new-inflexion}
  \left. \frac{\dd^2 \mu}{\dd m^2} \right|_{\sigma} &= 0 \;,
\end{align}
where 
\begin{align}
  \frac{\mu}{\Lambda} ( \sigma , x )  &= \frac{\sigma}{\eta x} \; , \\
  \frac{m}{\Lambda} ( \sigma , x ) &= \eta^{-1} \; .
\end{align}

At fixed $\sigma$, we can compute the total differentials: 
\begin{align}
  \label{eq:app-d-mu-over-lambda}
  \dd \left( \frac{\mu}{\Lambda} \right)&= 
            - \frac{\sigma}{x\eta^2} \dd \eta 
            - \frac{\sigma}{\eta x^2} \dd x \; , \\
  \label{eq:app-d-m-over-lambda}
  \dd \left( \frac{m}{\Lambda} \right)  &= - \frac{\dd\eta}{\eta^2} \;. 
\end{align}

From \refeq{eq:app-sys-CEP-new-MFE}, and using \refeq{eq:app-d-mu-over-lambda}
and \refeq{eq:app-d-m-over-lambda} we have:
\begin{equation}
  \label{eq:app-sys-CEP-reex-der}
 \eta F'(\eta)   + x Z_x(\sigma,x) = 0  \;,
\end{equation}
where $F'$ is the derivative of $F$ with respect to $\eta$, and $Z_x$ is the 
partial derivative of $Z$ with respect to $x$.

In the same fashion, we can write:
\begin{align}
  0 &= \left.\frac{\dd^2 \mu}{\dd m^2}\right|_\sigma \nonumber \\
    &= \frac{\dd}{\dd m} 
            \left[ - \frac{\sigma}{\eta x} 
                          \left( 
                                  \eta^{-1} + x^{-1} \frac{\dd x}{\dd \eta} 
                          \right) 
                  (-\eta^2)\right] \;.
\end{align}

After some manipulations, and using \refeq{eq:app-sys-CEP-reex-der}, one finds:
\begin{equation}
\eta^2 F''(\eta) - x^2 Z_{xx}(\sigma,x) - 2x Z_{x}(\sigma,x) = 0\;,
\end{equation}
where $Z_{xx}$ is the second partial derivative of $Z$ with respect to $x$.

The system to solve is now:
\begin{align}
  \label{eq:app-sys-fin-CEP-1}
  F(\eta) - Z(\sigma , x) &= 0 \; ,  \\
  \label{eq:app-sys-fin-CEP-2}
  \eta F'(\eta)  + x Z_{x}(\sigma,x) &= 0 \;, \\
  \label{eq:app-sys-fin-CEP-3}
  \eta^2 F''(\eta) - x^2 Z_{xx}(\sigma,x) - 2x Z_{x}(\sigma,x) &= 0\;. 
\end{align}

With the correct initialization, in particular if the initialization for 
$\eta$ is already the solution of \refeq{eq:app-inverse-of-F-eta-solution} 
for the initial values of $x$ and $\sigma$, a simple root finding algorithm 
can compute the solution in a few millisecond (this algorithm is about a 
hundred times faster than the usual algorithm).

Depending on $a$ and $b$ \refeq{eq:app-definition-of-coef-a-and-b},
the CEP may disappear. In that case, one should not try to solve the
system given by
\refeqs{eq:app-sys-fin-CEP-1}{eq:app-sys-fin-CEP-2}{eq:app-sys-fin-CEP-3}. To
detect if the CEP exists, it is always equivalent to have a metastable
solution at zero temperature. Hence, by solving $\dd^2 \mu / \dd^2 m =
0$, and looking at the value of $\dd \mu / \dd m$ at this point, one
can very efficiently (also a few millisecond) determine if the CEP
exists ($\dd \mu / \dd m > 0$) or not.

\section{Analytical derivation of the sigma-meson mass sensitivity}
\label{app:msigma-sensitivity}

To compute the sensitivity in the case of the sigma-meson mass:
\begin{equation}
  m_\sigma^2 = 4m^2 + m_\pi^2 \;,
\end{equation}
one can use the previous change of variable $m = \Lambda x$ and then:
\begin{equation}
m_\sigma^2 = 4 \Lambda^2 x^2 + m_\pi^2 \;.
\end{equation}
If $x$ is solution of \refeq{eq:app-decomp-sys-1}, then $\Lambda$ is
given by \refeq{eq:app-decomp-sys-3}, and the sigma meson mass reads:
\begin{equation}
m_\sigma = \sqrt{\frac{4}{N_c} \frac{f_\pi^2}{i_2} + m_\pi^2} \;.
\end{equation}

 With $c^3 = -\qbarq$ we have the differential:
 \begin{equation}
 d i_2 = \frac{d i_2}{dx} dx
 = - \frac{d i_2}{dx} \frac{1}{G'_\alpha(x)} d (f_\pi^3 / c^3) .
 \end{equation}
It is straightforward to compute $\dd m_\sigma^2 = 2m_\sigma \dd m_\sigma$ and
finally the needed partial derivatives. We found:
 \begin{align}
  \label{eq:dmp-ms}
  \frac{\partial m_\sigma}{\partial m_\pi} &=\frac{m_\pi}{m_\sigma}\;,\\
  \label{eq:dfp-ms}
  \frac{\partial m_\sigma}{\partial f_\pi} &= \frac{2 f_\pi}{m_\sigma N_c i_2} \left( 2 + \frac{3
      \alpha}{i_2} \frac{\dd i_2}{\dd x} \frac{1}{G'_\alpha} \right)\;,\\
  \label{eq:dc-ms}
\frac{\partial m_\sigma}{\partial  c} &= -6 \frac{f_\pi^2}{m_\sigma N_c
  i_2^2}\frac{\alpha}{c}\frac{\dd i_2}{\dd x}
\frac{1}{G'_\alpha} \;.
\end{align}
With the value of the inputs in the manuscript one finds: $ \dd m_\sigma =
0.21 \dd m_\pi + 34 \dd f_\pi - 8.2 \dd c$ so, for a vanishing relative
dispersion, $\Sigma(m_\sigma) = 6.42$. The value we obtained with the
Monte-Carlo, \reftab{tab:sensitivity}, is $\Sigma(m_\sigma) = 6.41$. 

The (small) difficulty here comes from the implicit equation for $x$. In
this particular case, all the quantities that appear in the inverse
problem only depend on the solution of \refeq{eq:app-decomp-sys-1}
which only depends on the quantity $\alpha = f_\pi^3 / c^3$, and then
one can access the sensitivity of the in-vacuum predictions quite
easily. With more realistic models, the inverse problem will not be
equivalent anymore to a one dimensional equation, and the Monte Carlo
becomes a better alternative.

\section{Analytical calculation of composed probability distributions}
\label{app:ana-prob-dist}

To check if the Monte-Carlo results are correct, one can compare them
to the theoretical probability distributions when possible. 

In the one variable case, the composition of two probability
distributions is as follow. Let's call $X$ and $Y$ two random
variables, with $X$ following its probability distribution $\rho_X$, and
$Y = f(X)$ ({\it i.e.} $Y$ is a function of the random variable
$X$). Let's call $x$ and $y$ the realization of the random variables
$X$ and $Y$ through their corresponding probability distributions.

If the function $f$ is monotonic and increasing, then the probability
of finding $x$ between $x_1$ and $x_2$ ($x_1 < x_2$) is equal to the
probability of finding $y$ between $y_1 = f(x_1)$ and $y_2 = f(x_2)$:
\begin{equation}
  \mrm{P} ( x_1 \leq x \leq x_2 ) = \mrm{P} \left( y_1 = f(x_1 ) \leq y \leq y_2 = f(x_2) \right) \;.
\end{equation}

By definition of the probability distribution we have:
\begin{align}
  \mrm{P} ( x_1 \leq x \leq x_2 ) &= \int_{x_1}^{x_2} \rho_X(x) \dd x \; ; \\
  \mrm{P} ( y_1 \leq y \leq y_2 ) &= \int_{f(x_1)}^{f(x_2)} \rho_Y(x) \dd y \; .
\end{align}

Then $\rho_X$ can be expressed as:
\begin{equation}
  \rho_X ( x ) = \left( \rho_Y \circ f \right) (x) f'(x) \;,
\end{equation}
which imply:
\begin{equation}
\label{eq:app-rho-1v}
  \rho_Y ( y ) = 
           \left(\rho_X \circ f^{-1}\right) (y) 
           \left[ \left(f' \circ f^{-1}\right)(y) \right]^{-1} \;.
\end{equation}

To illustrate the case of two variables, we give the result
for the density $\rho_\alpha$. We have, following the same treatment as
for the one variable case, and using the shortcuts $f$ for $f_\pi$, and $c$ for 
$\qbarq$:
\begin{align}
  \mrm{P} ( \alpha_1 \leq \alpha  \leq \alpha_2 ) 
        &=  \mrm{P} \left( \frac{f^3}{c} \in [\alpha_1 , \alpha_2]\right) 
               \nonumber \\
        &= \mrm{P} \left(  \frac{f^3}{c} \geq \alpha_1 \;\;\wedge\;\; 
                  \frac{f^3}{c}\leq \alpha_2) \right) \nonumber \\
        &= \mrm{P} \left( f \in \mathbb{R}^+ \;\; \wedge \;\;  
                      \frac{f^3}{\alpha_2} \leq x \leq  \frac{f^3}{\alpha_1} 
                     \right) \nonumber \\
        &= \int_{\mathbb{R}^+} \dd f 
             \int_{f^3/\alpha_2}^{f^3/\alpha_1} \dd c \; 
                     \rho_f(f)\rho_c(c) \;.
\end{align}

This probability can be re-expressed using the probability distribution
$\rho_\alpha$:
\begin{align}
  \mrm{P} ( \alpha_1 \leq \alpha  \leq \alpha_2 ) &= 
        \int_{\alpha_1}^{\alpha_2} \rho_\alpha(\alpha) \dd \alpha \;,
\end{align}
and then we find an expression for  $\rho_\alpha$:
\begin{equation}
  \rho_\alpha(\alpha) = 
             \frac{\dd}{\dd \alpha}  
                 \int_{\mathbb{R}^+} \dd f 
                     \int_{f^3/\alpha}^{f^3/\alpha^*} \dd c \; 
                          \rho_f(f)\rho_c(c) \:,
\end{equation}
where $\alpha^*$ is any constant. Finally $\rho_\alpha$ is found to read:
\begin{equation}
  \label{eq:app-rho-alpha-theo}
  \rho_\alpha(\alpha) = \frac{1}{\alpha^2}  
                         \int_{\mathbb{R}^+} \dd f \; f^3  \rho_f(f)\rho_c(f^3/\alpha) \:.
\end{equation}

\section{Kernel density approximation}
\label{app:Kernel density approximation}

The statistical technique we used allowed us to draw some scatter
plots for the CEP coordinates prediction. It is also interesting to
have access to the probability distribution of the CEP, \idest the
density of points, in the $(T - \mu)$ plane. In order to reconstruct
the density from the data, a possibility is to use the Kernel Density
Estimate (KDE) with Gaussian kernels.

Following Eqs.(4), (5), (6) and (7) of \ccite{Hwang324744} we can
reconstruct the density $\rho(T_{CEP}, \mu_{CEP})$. This algorithm
normalizes the density to get a probability distribution \\
($\int \rho(T,\mu)\, dT\, d\mu = 1$) hence its dimension is GeV$^{-2}$. To do
this:

First we compute the covariance matrix and transform the data to
obtain a set of data with zero mean value and unity standard deviation
-- the so-called sphered data. One has simply to apply to data the
matrix $S^{1/2}$ where $S$ is the covariance matrix.

Then each sphered data point is replaced by a Gaussian with
a variance chosen such as its standard deviation is large
enough to overlap with other data points but small enough not to
create a long tail that does not exist in data. This is the smoothing
procedure control by the smoothing parameter $h$ of Hwang. We check
that with the parameter $h$ given in the paper, we are able to
reconstruct very well a two dimensional Gaussian distribution with as
low as a hundred point (it is the ``easy case'' for this algorithm)
but also to reconstruct quite well a 2D uniform distribution also with
a hundred points (the difficult case).

Finally the reconstructed density is the sum of the smoothed sphered
data where the matrix $S^{-1/2}$ is applied to get back to the original
data.

\end{appendix}


\begin{thebibliography}{10}

\bibitem{Asakawa:1989bq}
M.~Asakawa and K.~Yazaki.
\newblock {Chiral Restoration at Finite Density and Temperature}.
\newblock {\em Nucl.Phys.}, A504:668--684, 1989.

\bibitem{Abelev:2009bw}
B.I. Abelev et~al.
\newblock {Identified particle production, azimuthal anisotropy, and
  interferometry measurements in Au+Au collisions at s(NN)**(1/2) = 9.2- GeV}.
\newblock {\em Phys.Rev.}, C81:024911, 2010.

\bibitem{Aggarwal:2010cw}
M.M. Aggarwal et~al.
\newblock {An Experimental Exploration of the QCD Phase Diagram: The Search for
  the Critical Point and the Onset of De-confinement}.
\newblock 2010.

\bibitem{Tarnowsky:2011vk}
Terence~J. Tarnowsky.
\newblock {Searching for the QCD Critical Point Using Particle Ratio
  Fluctuations and Higher Moments of Multiplicity Distributions}.
\newblock {\em J.Phys.}, G38:124054, 2011.

\bibitem{Lacey:2014wqa}
Roy~A. Lacey.
\newblock {Indications for a Critical End Point in the Phase Diagram for Hot
  and Dense Nuclear Matter}.
\newblock {\em Phys.Rev.Lett.}, 114(14):142301, 2015.

\bibitem{Akiba:2015jwa}
Yasuyuki Akiba, Aaron Angerami, Helen Caines, Anthony Frawley, Ulrich Heinz,
  et~al.
\newblock {The Hot QCD White Paper: Exploring the Phases of QCD at RHIC and the
  LHC}.
\newblock 2015.

\bibitem{CPOD}
{\em 8th International Workshop on Critical Point and Onset of Deconfinement},
  2013.

\bibitem{Luo:2009sx}
Xiaofeng Luo, Ming Shao, Cheng Li, and Hongfang Chen.
\newblock {Signature of QCD critical point: Anomalous transverse velocity
  dependence of antiproton-proton ratio}.
\newblock {\em Phys.Lett.}, B673:268--271, 2009.

\bibitem{Gavai:2011sr}
Rajiv~V. Gavai.
\newblock {QCD Critical Point: Synergy of Lattice and Experiments}.
\newblock {\em Acta Phys.Polon.}, B43:723--730, 2012.

\bibitem{Gazdzicki:2011fx}
Marek Gazdzicki.
\newblock {NA49/NA61: results and plans on beam energy and system size scan at
  the CERN SPS}.
\newblock {\em J.Phys.}, G38:124024, 2011.

\bibitem{NICA}
D.~Blaschke et~al.
\newblock {\em Searching for for a QCD Mixed Phase at the Nuclotron-Based Ion
  Collider Facility (NICA White Paper)}.
\newblock Dubna, 2013.

\bibitem{Fodor:2004nz}
Z.~Fodor and S.D. Katz.
\newblock {Critical point of QCD at finite T and mu, lattice results for
  physical quark masses}.
\newblock {\em JHEP}, 0404:050, 2004.

\bibitem{Borsanyi:2010cj}
Szabolcs Borsanyi, Gergely Endrodi, Zoltan Fodor, Antal Jakovac, Sandor~D.
  Katz, et~al.
\newblock {The QCD equation of state with dynamical quarks}.
\newblock {\em JHEP}, 1011:077, 2010.

\bibitem{Bazavov:2011nk}
A.~Bazavov, T.~Bhattacharya, M.~Cheng, C.~DeTar, H.T. Ding, et~al.
\newblock {The chiral and deconfinement aspects of the QCD transition}.
\newblock {\em Phys.Rev.}, D85:054503, 2012.

\bibitem{Fukushima:2004}
K.~Fukushima.
\newblock Chiral effective model with the polyakov loop.
\newblock {\em Physics Letters B}, 591:277--284, 2004.

\bibitem{Ratti:2006}
C.~Ratti, M.A. Thaler, and W.~Weise.
\newblock Phases of {QCD}: Lattice thermodynamics and a field theoretical
  model.
\newblock {\em Physical Review D}, 73(014019), 2006.

\bibitem{Costa:2007ie}
Pedro Costa, C.A. de~Sousa, M.C. Ruivo, and Yu.L. Kalinovsky.
\newblock {The QCD critical end point in the SU(3) Nambu-Jona-Lasinio model}.
\newblock {\em Phys.Lett.}, B647:431--435, 2007.

\bibitem{Fukushima:2008b}
K.~Fukushima.
\newblock Phase diagrams in the three-flavor nambu jona-lasinio model with the
  polyakov loop.
\newblock {\em Physical Review D}, 77(114028), 2008.

\bibitem{Kashiwa:2008a}
K.~Kashiwa, H.~Kouno, M.~Matsuzaki, and M.~Yahiro.
\newblock Critical endpoint in the polyakov loop extended njl model.
\newblock {\em Physics Letters B}, 662:26--32, 2008.

\bibitem{Rossner:2008}
S.~Rossner, T.~Hell, C.~Ratti, and W.~Weise.
\newblock The chiral and deconfinement crossover transitions: Pnjl model beyond
  mean field.
\newblock {\em Nuclear Physics A}, 814:118--143, 2008.

\bibitem{Costa:2008gr}
Pedro Costa, C.A. de~Sousa, M.C. Ruivo, and H.~Hansen.
\newblock {The QCD critical end point in the PNJL model}.
\newblock {\em Europhys.Lett.}, 86:31001, 2009.

\bibitem{PNJLPCHH1}
P.~Costa, H.~Hansen, M.~C. Ruivo, and C.~A. de~Sousa.
\newblock How parameters and regularization affect the
  polyakov–nambu–jona-lasinio model phase diagram and thermodynamic
  quantities.
\newblock {\em Phys. Rev. D}, 81:016007, Jan 2010.

\bibitem{Costa2010}
Pedro Costa, M.C. Ruivo, C.A. de~Sousa, and H.~Hansen.
\newblock {Phase diagram and critical properties within an effective model of
  QCD: the Nambu-Jona-Lasinio model coupled to the Polyakov loop}.
\newblock {\em Symmetry}, 2:1338--1374, 2010.

\bibitem{Schaefer:2007pw}
Bernd-Jochen Schaefer, Jan~M. Pawlowski, and Jochen Wambach.
\newblock {The Phase Structure of the Polyakov--Quark-Meson Model}.
\newblock {\em Phys.Rev.}, D76:074023, 2007.

\bibitem{Herbst:2010rf}
Tina~Katharina Herbst, Jan~M. Pawlowski, and Bernd-Jochen Schaefer.
\newblock {The phase structure of the Polyakov--quark-meson model beyond mean
  field}.
\newblock {\em Phys.Lett.}, B696:58--67, 2011.

\bibitem{Fukushima:2008wg}
Kenji Fukushima.
\newblock {Phase diagrams in the three-flavor Nambu-Jona-Lasinio model with the
  Polyakov loop}.
\newblock {\em Phys.Rev.}, D77:114028, 2008.

\bibitem{Bratovic:2012qs}
Nino~M. Bratovic, Tetsuo Hatsuda, and Wolfram Weise.
\newblock {Role of Vector Interaction and Axial Anomaly in the PNJL Modeling of
  the QCD Phase Diagram}.
\newblock {\em Phys.Lett.}, B719:131--135, 2013.

\bibitem{Contrera:2012wj}
G.A. Contrera, A.G. Grunfeld, and D.B. Blaschke.
\newblock {Phase diagrams in nonlocal PNJL models constrained by Lattice QCD
  results}.
\newblock 2012.

\bibitem{Hell:2012da}
Thomas Hell, Kouji Kashiwa, and Wolfram Weise.
\newblock {Impact of vector-current interactions on the QCD phase diagram}.
\newblock {\em J.Mod.Phys.}, 4:644--650, 2013.

\bibitem{Kaczmarek:2011zz}
O.~Kaczmarek, F.~Karsch, E.~Laermann, C.~Miao, S.~Mukherjee, et~al.
\newblock {Phase boundary for the chiral transition in (2+1) -flavor QCD at
  small values of the chemical potential}.
\newblock {\em Phys.Rev.}, D83:014504, 2011.

\bibitem{Blaschke:2013ana}
David Blaschke, David~E. Alvarez-Castillo, and Sanjin Benic.
\newblock {Mass-radius constraints for compact stars and a critical endpoint}.
\newblock {\em PoS}, CPOD2013:063, 2013.

\bibitem{Tarantola}
Albert Tarantola.
\newblock {\em {I}nverse {P}roblem {T}heory and {M}ethods for {M}odel
  {P}arameter {E}stimation}.
\newblock Society for Industrial and Applied Mathematics, 2005.

\bibitem{Ramm:1997}
A.G. Ramm.
\newblock {\em Inverse problems, tomography, and image processing}.
\newblock Springer, 1997.

\bibitem{Dobaczewski1}
J.~Dobaczewski, W.~Nazarewicz, and P.-G. Reinhard.
\newblock {Error Estimates of Theoretical Models: a Guide}.
\newblock {\em J.Phys.}, G41:074001, 2014.

\bibitem{Dobaczewski2}
J.~Toivanen, J.~Dobaczewski, M.~Kortelainen, and K.~Mizuyama.
\newblock Error analysis of nuclear mass fits.
\newblock {\em Phys. Rev. C}, 78:034306, Sep 2008.

\bibitem{Reinhard:2010}
P.G. Reinhard and W.~Nazarewicz.
\newblock Information content of a new observable: The case of the nuclear
  neutron skin.
\newblock {\em Phys.Rev.C}, 81(051303(R)), 2010.

\bibitem{Fattoyev:2011}
F.J. Fattoyev and J.~Piekarewicz.
\newblock Accurate calibration of relativistic mean-field models: Correlating
  observables and providing meaningful theoretical uncertainties.
\newblock {\em Phys.Rev.C}, 84(064302), 2010.

\bibitem{Kortelainen2010}
M.~Kortelainen, T.~Lesinski, J.~More, W.~Nazarewicz, J.~Sarich, et~al.
\newblock {Nuclear Energy Density Optimization}.
\newblock {\em Phys.Rev.}, C82:024313, 2010.

\bibitem{FelipeCucker2002}
Mario~Wschebor Felipe~Cucker.
\newblock On the expected condition number of linear programming problems.
\newblock {\em NumerischeMathematik}, 2002.

\bibitem{Demmel:1987}
J.W. Demmel.
\newblock On condition numbers and the distance of the nearest ill-pose
  problem.
\newblock {\em Numerische Mathematik}, 51:251, 1987.

\bibitem{NJLrev_klevansky}
S.~P. Klevansky.
\newblock The {N}ambu—{J}ona-{L}asinio model of quantum chromodynamics.
\newblock {\em Rev. Mod. Phys.}, 64:649--708, Jul 1992.

\bibitem{NJLrev_weise}
U.~Vogl and W.~Weise.
\newblock The {N}ambu and {J}ona-{L}asinio model: {I}ts implications for
  {H}adrons and {N}uclei.
\newblock {\em Progress in Particle and Nuclear Physics}, 27(0):195 -- 272,
  1991.

\bibitem{NJLrev_buballa}
Michael Buballa.
\newblock {NJL}-model analysis of dense quark matter.
\newblock {\em Physics Reports}, 407(4–6):205 -- 376, 2005.

\bibitem{NJLrev_hatsuda}
Tetsuo Hatsuda and Teiji Kunihiro.
\newblock {QCD} phenomenology based on a chiral effective lagrangian.
\newblock {\em Physics Reports}, 247(5–6):221 -- 367, 1994.

\bibitem{Dosch1998173}
H.G. Dosch and S.~Narison.
\newblock Direct extraction of the chiral quark condensate and bounds on the
  light quark masses.
\newblock {\em Physics Letters B}, 417(1–2):173 -- 176, 1998.

\bibitem{BordesJHEP2010}
J.~Bordes, C.A. Dominguez, P.~Moodley, J.~Peñarrocha, and K.~Schilcher.
\newblock Chiral corrections to the su(2)xsu(2) gell-mann-oakes-renner
  relation.
\newblock {\em Journal of High Energy Physics}, 2010(5), 2010.

\bibitem{Aoki:2013ldr}
Sinya Aoki, Yasumichi Aoki, Claude Bernard, Tom Blum, Gilberto Colangelo,
  et~al.
\newblock {Review of lattice results concerning low energy particle physics}.
\newblock {\em arXiv}, 1310.8555, 2013.

\bibitem{Lourenco2012}
O.~Lourenco, M.~Dutra, T.~Frederico, A.~Delfino, and M.~Malheiro.
\newblock {Vector interaction strength in Polyakov-Nambu-Jona-Lasinio models
  from hadron-quark phase diagrams}.
\newblock {\em Phys.Rev.}, D85:097504, 2012.

\bibitem{Oertel2000}
Micaela Oertel.
\newblock {\em Investigation of meson loop effects in the Nambu-Jona-Lasinio
  model}.
\newblock doctoral dissertation ar{X}iv: hep-ph/0012224, 2000.

\bibitem{Celenza2000}
L.S. Celenza, Shun-fu Gao, Bo~Huang, Huangsheng Wang, and C.M. Shakin.
\newblock {Covariant confinement model for the calculation of the properties of
  scalar mesons}.
\newblock {\em Phys.Rev.}, C61:035201, 2000.

\bibitem{Celenza2001}
L.~S. Celenza, Huangsheng Wang, and C.~M. Shakin.
\newblock {Application of a generalized Nambu-Jona-Lasinio model to the
  calculation of the properties of scalar mesons and nuclear matter}.
\newblock {\em Phys.Rev.}, C63:025209, 2001.

\bibitem{Hwang324744}
Jenq-Neng Hwang, S.-R. Lay, and A~Lippman.
\newblock Nonparametric multivariate density estimation: a comparative study.
\newblock {\em Signal Processing, IEEE Transactions on}, 42(10):2795--2810, Oct
  1994.

\end{thebibliography}
\end{document}